\documentclass[aps,preprint,showpacs,superscriptaddress,groupedaddress]{revtex4-1}  

\usepackage{graphicx}  
\usepackage{dcolumn}   
\usepackage{bm}        
\usepackage{amssymb}   
\usepackage{amsmath}   
\usepackage{listings}
\usepackage{array}
\usepackage{graphicx}
\usepackage{lineno}
\usepackage{dcolumn}
\usepackage{color}
\usepackage{overpic}
\usepackage{multirow}
\usepackage{hyperref}
\usepackage{cleveref}

\newcommand{\pt}{p_{\text{T}}}
\newcommand{\SM}{{\text{SM}}}

\begin{document}


\title{$t$-channel Higgs production and constraints on off-shell Higgs couplings}
\author{Li-Gang Xia \\ School of Physics, Nanjing University, Nanjing, China, 210000}

\begin{abstract}
    This note is to record an unsucessful idea. I tried hard to persuade myself to write this note as I have spent a significant fraction of time during this summer. For Higgs resonant production and decaying in the $s$ channel, $i \to H \to f$, the cross section is proportional to $y_i^2y_f^2/\Gamma_H$. Thus there is a multi-solution ambiguity on the couplings due to uncertain knowledge on Higgs width. In this work, we study the process $g+q \to g+q+l^++l^-$ with the $t$-channel Higgs boson contribution included. As Higgs boson is not on-shell, this process can be used to measure the Higgs couplings without the ambiguity.
    Based on the performance of the ATLAS detector in Run~II, the coupling product $y_gy_Z$ can be potentially constrained to be less than 18 times the Standard-Model value for a $p$-$p$ dataset of 140~fb$^{-1}$ at $\sqrt{s}=13$~TeV. Assuming the equality of on-shell couplings and off-shell couplings and combining the on-shell measurements, the width of the Higgs boson is constrained to be less than 1.3~GeV, which is similar to that in the direct measurement.
    In addition, the interference effect is also analyzed.

\end{abstract}
\maketitle 

\section{Introduction}
After the discovery of the Higgs boson~\cite{higgs_observation_atlas,higgs_observation_cms}, we have been in the era of precision measurements of the Higgs boson properties. For the Higgs production in a virtual process $g \to g + H$, supposing the incoming and outgoing four momenta are $p_i$ and $p_o$, the squared momentum transfer $q^2$ is
\begin{equation}
    q^2 = (p_i-p_o)^2 = (E_i-E_o)^2 - (\vec{p}_i-\vec{p}_o)^2 = -2(E_iE_o-\vec{p}_i\cdot\vec{p}_o) \leq 0 \:,
\end{equation}
which is non-positive. Hence the Higgs boson will never be on-shell and the Higgs propagator, $1/(q^2-m_H^2+im_H\Gamma_H)$, has then little dependence upon its width in the Standard Model (SM). We care about off-shell Higgs couplings because we would assume they are the same as the on-shell ones and measure the Higgs width using the so-called ``off-shell'' method~\cite{higgswidth0,higgswidth1}. 

In this paper, we propose to measure the off-shell Higgs couplings using the $t$-channel Higgs production processes like $g+q \to g+q+ l^+l^-$ ($l=e,\mu$).

\section{Background analysis and Inteference effect}
The dominant background events are those with two leptons plus one or more jets. Its cross section is huge ($~10^2$~pb) while the signal cross section is tiny ($10^{-5}$~pb). The background can be classified into two classes, incoherent and coherent parts. The coherent background would interfere with the signal as they share exactly the same initial and final states. Unfortunately, as we will see in next section, the interference effect is destructive and I am NOT able to use the
interference effect to amplify the signal sensitivty. Though the interference effect is much larger than the signal strength itself, I am NOT able to find a phase space which is strongly sensitive to the interference effect, which would potentially indicate the existence of the $t$-channel Higgs. More details can be found in next section. 

Let us explain the observation above using the following simple model.
\begin{eqnarray}
    \sigma(\kappa) = && \sigma_{b_0} + \sigma_{b_1} + \kappa^2\sigma_{s} + 2\sqrt{\sigma_{b_1}\kappa^2\sigma_s}\cos\delta \\
    = && \sigma_b + \kappa^2\sigma_s + 2\kappa\sqrt{\sigma_b\sigma_s }\sqrt{f}\cos\delta \\
    = && \sigma_b + \sigma_s(\kappa) + \sigma_I(\kappa)
\end{eqnarray}
where $\kappa\equiv \frac{y_gy_Z}{y_g^{\SM}y_Z^{\SM}}$ is a coupling modifier; $\sigma_b$ denotes the total background cross section; $\sigma_s$ denotes the signal cross section; $f$ denotes the fraction of coherent background cross section, namely, $f\equiv \sigma_{b_1}/\sigma_b$; and $\delta$ denotes the relative phase angle.

In our case, if we do not select a specific phase space, we have
\begin{equation}
    \sigma_b >> \sigma_I(\kappa) \gtrsim \sigma_s(\kappa) \:,
\end{equation}
for a not too big $\kappa$ like $\lesssim50$.

My initial idea is to use the interference effect to probe the Higgs couplings instead of the signal itself (because it is too small). However I cannot find a good phase space so that the interference effect is relatively significant. It is because
\begin{equation}
    \frac{\sigma_I(\kappa)}{\sigma_b} << 1 \: , 
\end{equation}
for a not too big $\kappa$ like $\lesssim50$.

After many trials and errors, I finally find a phase space, where the $Z$ boson (the di-lepton system) is highly boosted ($\pt(ll)>200$~GeV).  In this region, the interference effect is minor and the signal is significant, namely, 

\begin{equation}
    \frac{\sigma_I(\kappa)}{\sigma_s(\kappa)} \lesssim 1 \: , 
\end{equation}
and
\begin{equation}
    \frac{\sigma_s(\kappa)}{\sigma_b} \lesssim 1 \: , 
\end{equation}
for a not too small $\kappa \gtrsim 10$. This is roughly the expected sensitivity for a detector like ATLAS and a $p$-$p$ dataset of 140~fb$^{-1}$ at $\sqrt{s}=13$~TeV.

\section{Interference effect at generator level}
First of all, let us define the fiducial region: $\pt(l)>5$~GeV, $\pt(ll)>180$~GeV, $|m(ll)-90|<50$~GeV, $|\eta(j)|<5$ and there is no explicit cut on $\pt(j)$ (because it is a $t$-channel process like the Vector-Boson-Fusion Higgs production, but the generator MadGraph actually has a threshold cut of around 10~GeV). 

The signal sample, background sample and the sample with the interference included are splitted into four parts according to $\pt(ll)$ by design, namely,
\begin{itemize}
    \item $\pt(ll)<180$~GeV: interference effect is significant compared to the signal, but the effect is tiny relative to the background size.
    \item $180<\pt(ll)<400$, $400<\pt(ll)<800$, $\pt(ll)>800$~GeV : interference effect is minor compared to the signal for a not small coupling modifier $\kappa$.
\end{itemize}

\begin{figure}[htbp]
     \centering
     \includegraphics[width=0.45\textwidth]{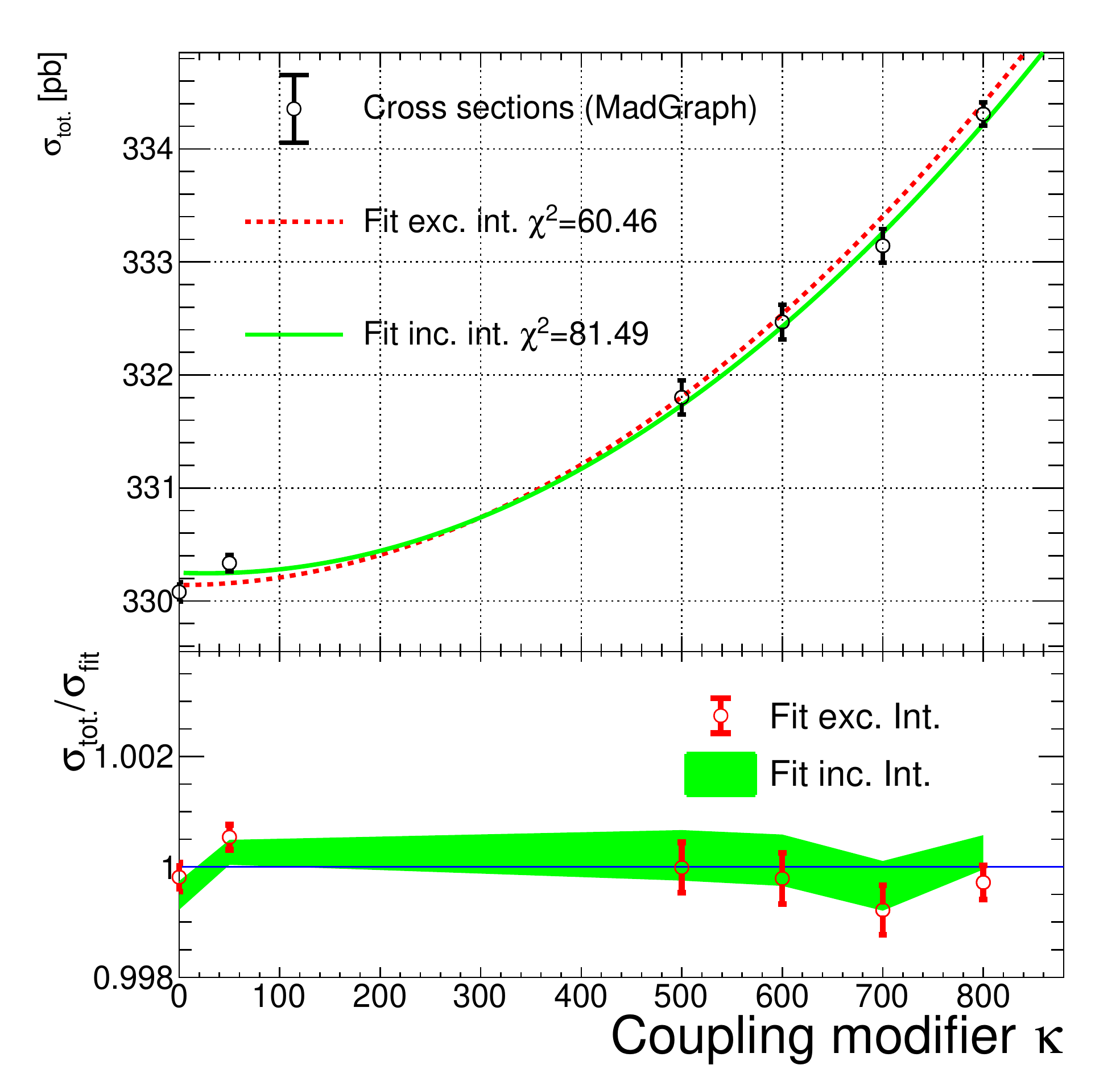}
     \includegraphics[width=0.45\textwidth]{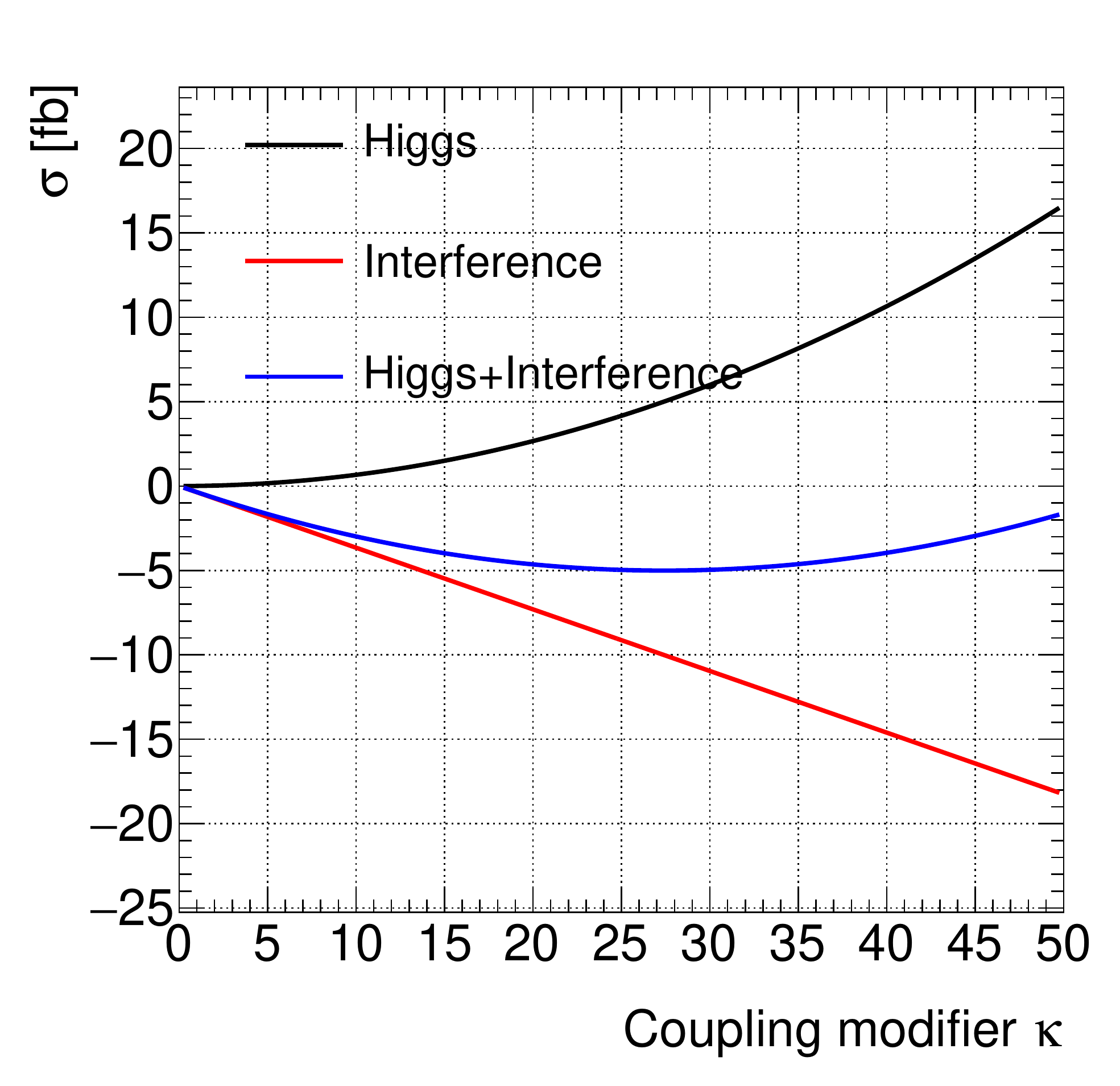}
     \caption{\label{fig:xs_ptll180}
     Total fiducial cross secrtion as a function of the coupling modifier $\kappa$ (Left) and the interference cross section (Right) for the region $\pt(ll)<180$~GeV. On the left plot, the green solid curve and red dashed curve show the fit results with or without including the interference term.
     }
 \end{figure}

\begin{figure}[htbp]
     \centering
     \includegraphics[width=0.45\textwidth]{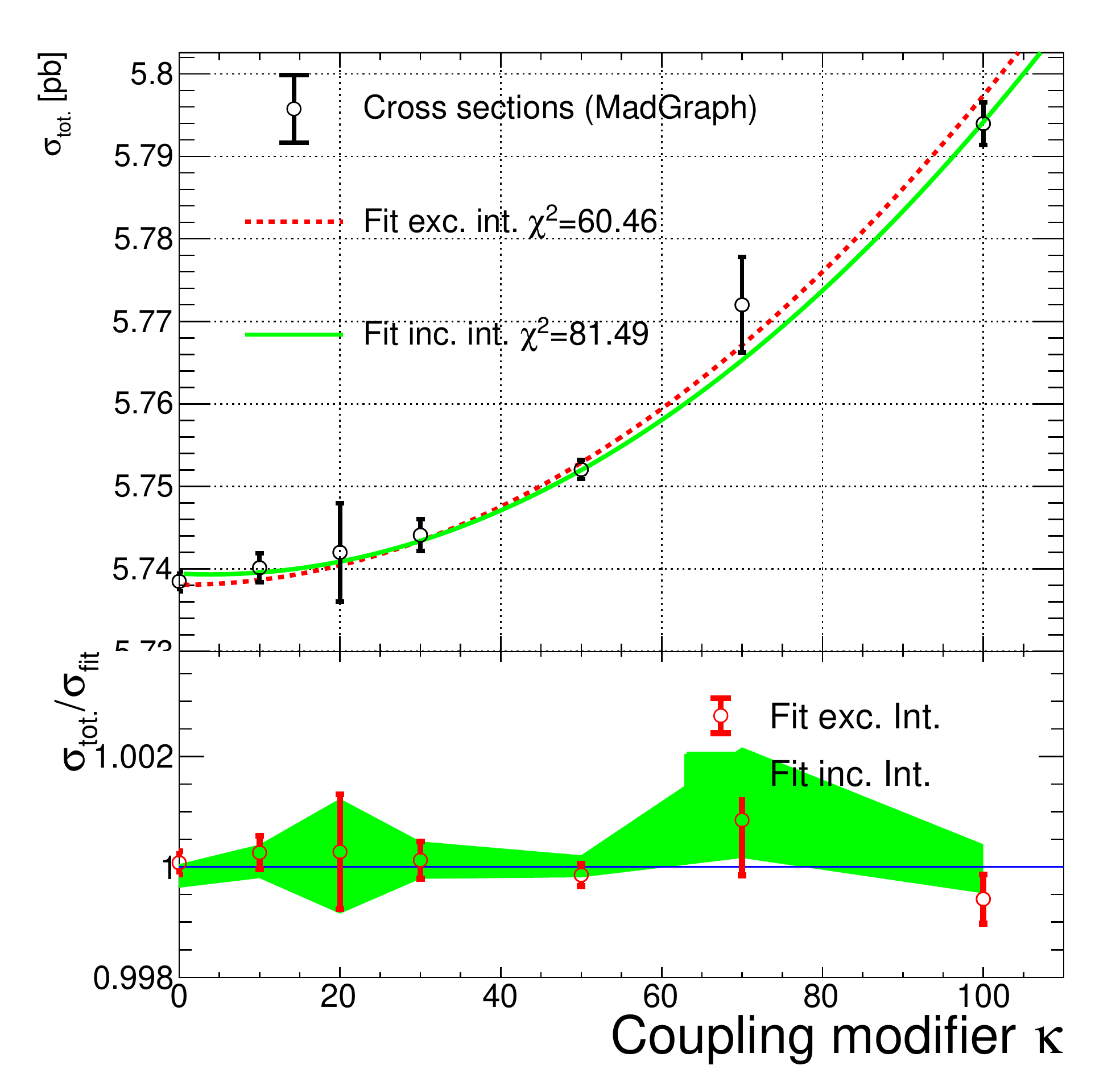}
     \includegraphics[width=0.45\textwidth]{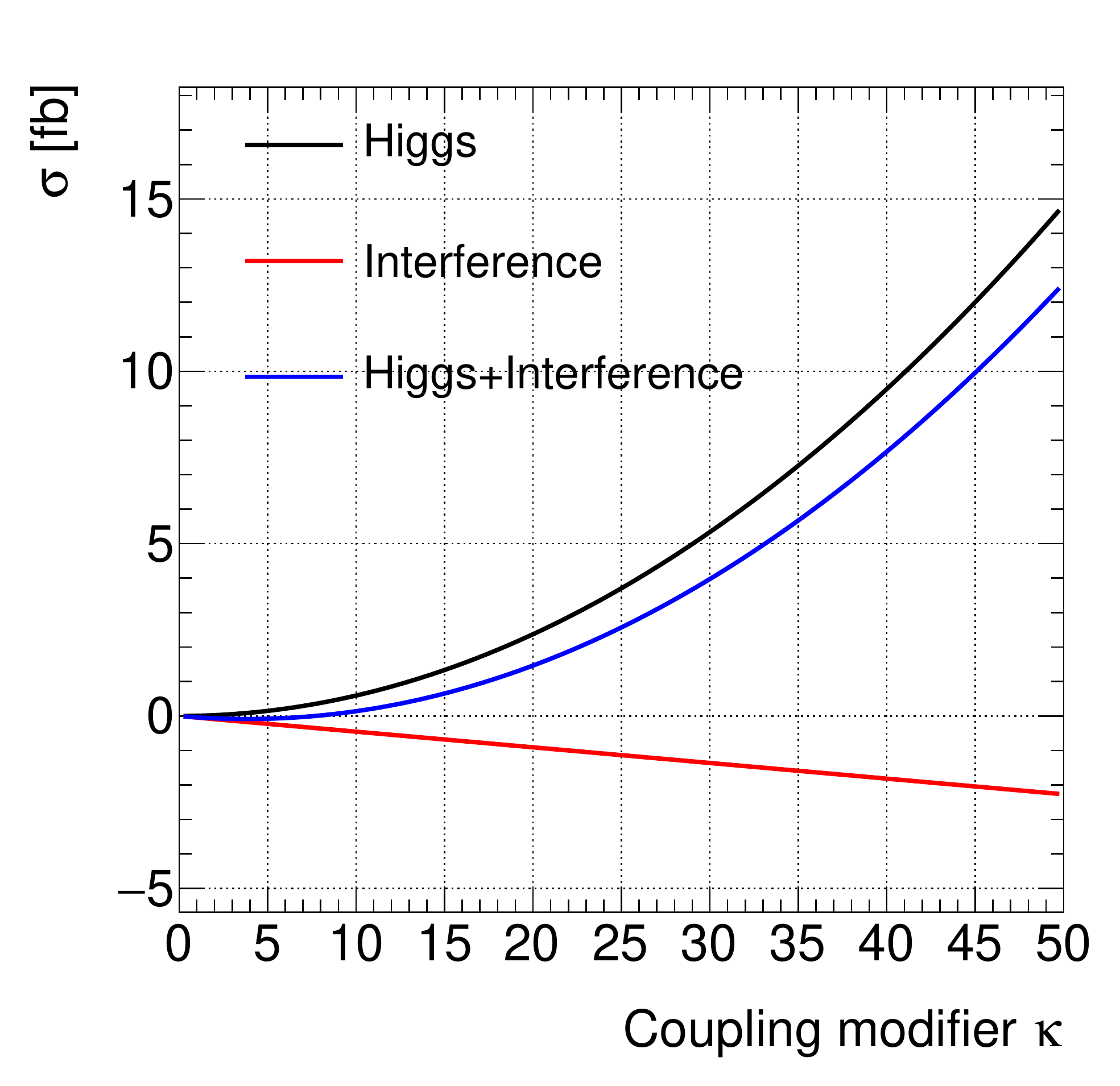}
     \caption{\label{fig:xs_ptll180to400}
     Total fiducial cross secrtion as a function of the coupling modifier $\kappa$ (Left) and the interference cross section (Right) for the region $180<\pt(ll)<400$~GeV. On the left plot, the green solid curve and red dashed curve show the fit results with or without including the interference term.
     }
 \end{figure}

\begin{figure}[htbp]
     \centering
     \includegraphics[width=0.45\textwidth]{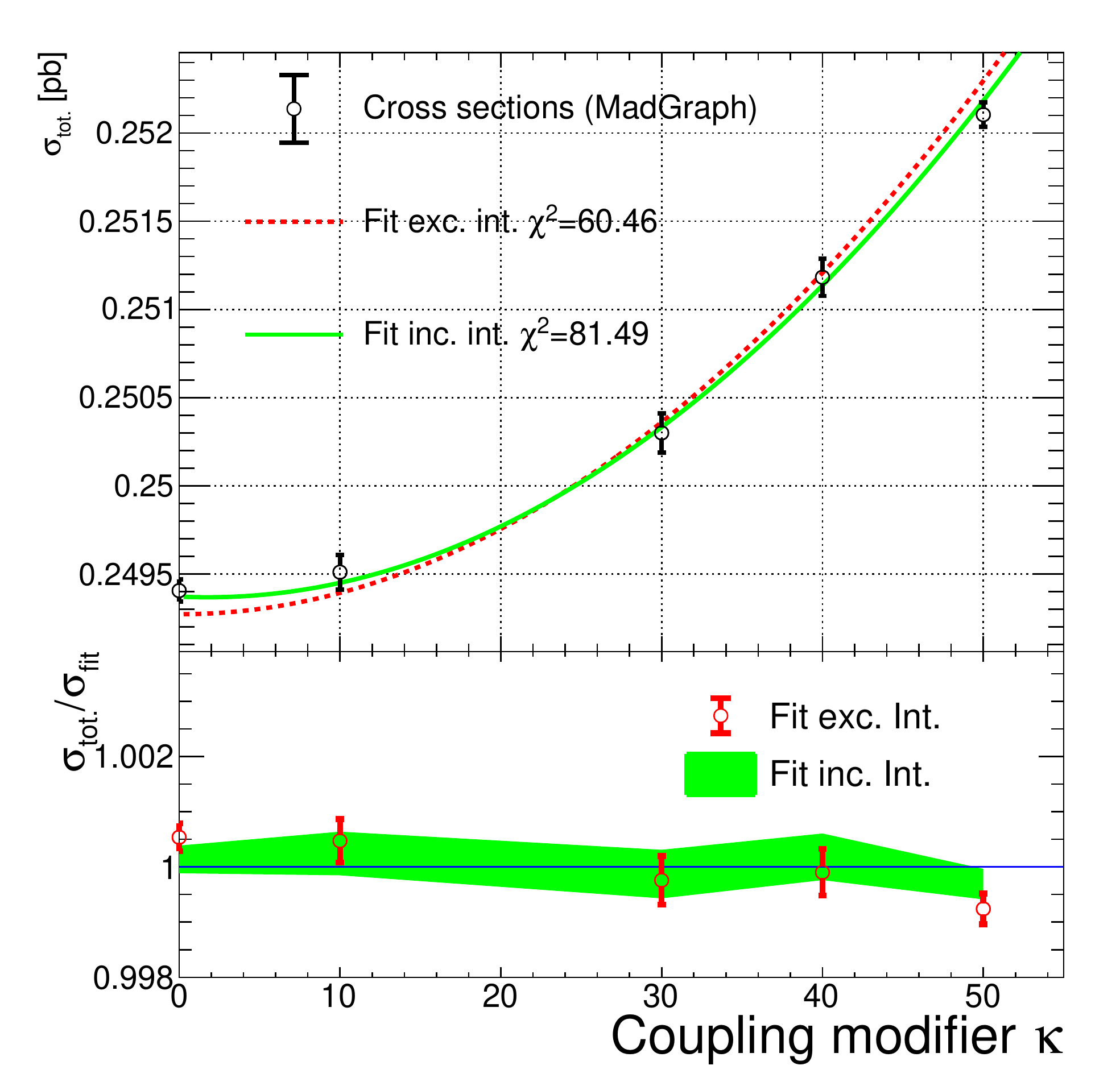}
     \includegraphics[width=0.45\textwidth]{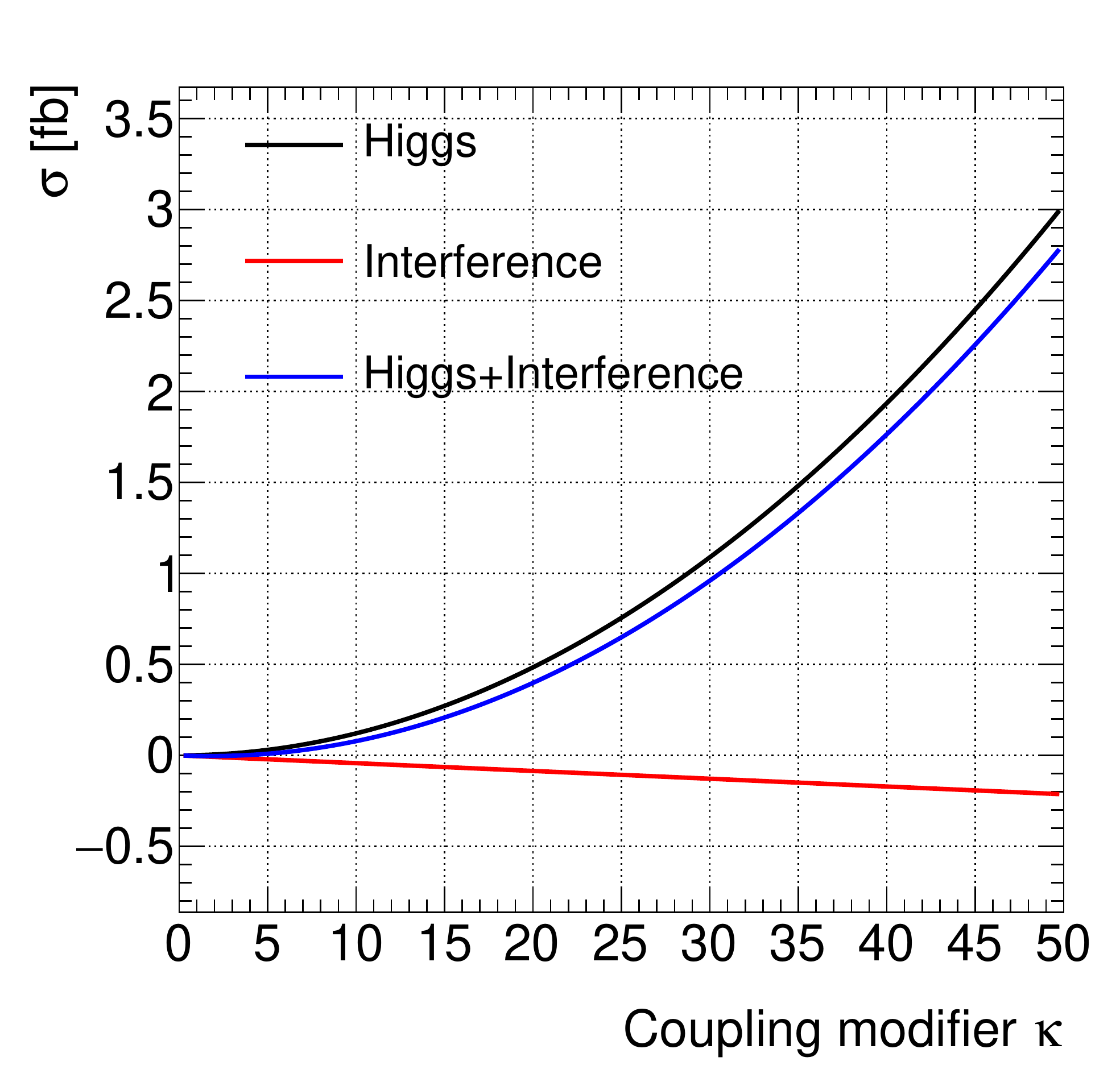}
     \caption{\label{fig:xs_ptll400to800}
     Total fiducial cross secrtion as a function of the coupling modifier $\kappa$ (Left) and the interference cross section (Right) for the region $400<\pt(ll)<800$~GeV. On the left plot, the green solid curve and red dashed curve show the fit results with or without including the interference term.
     }
 \end{figure}

\begin{figure}[htbp]
     \centering
     \includegraphics[width=0.45\textwidth]{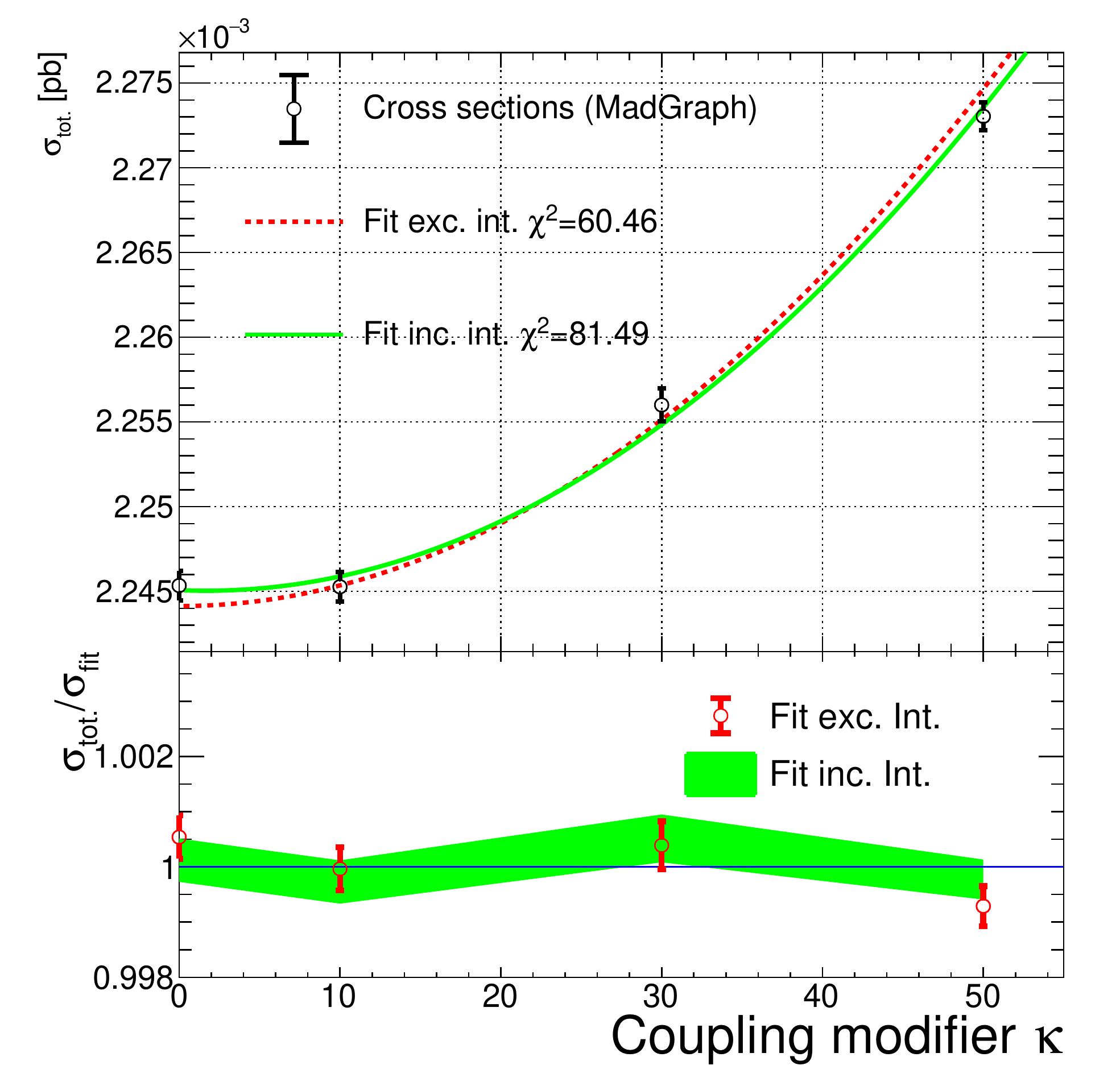}
     \includegraphics[width=0.45\textwidth]{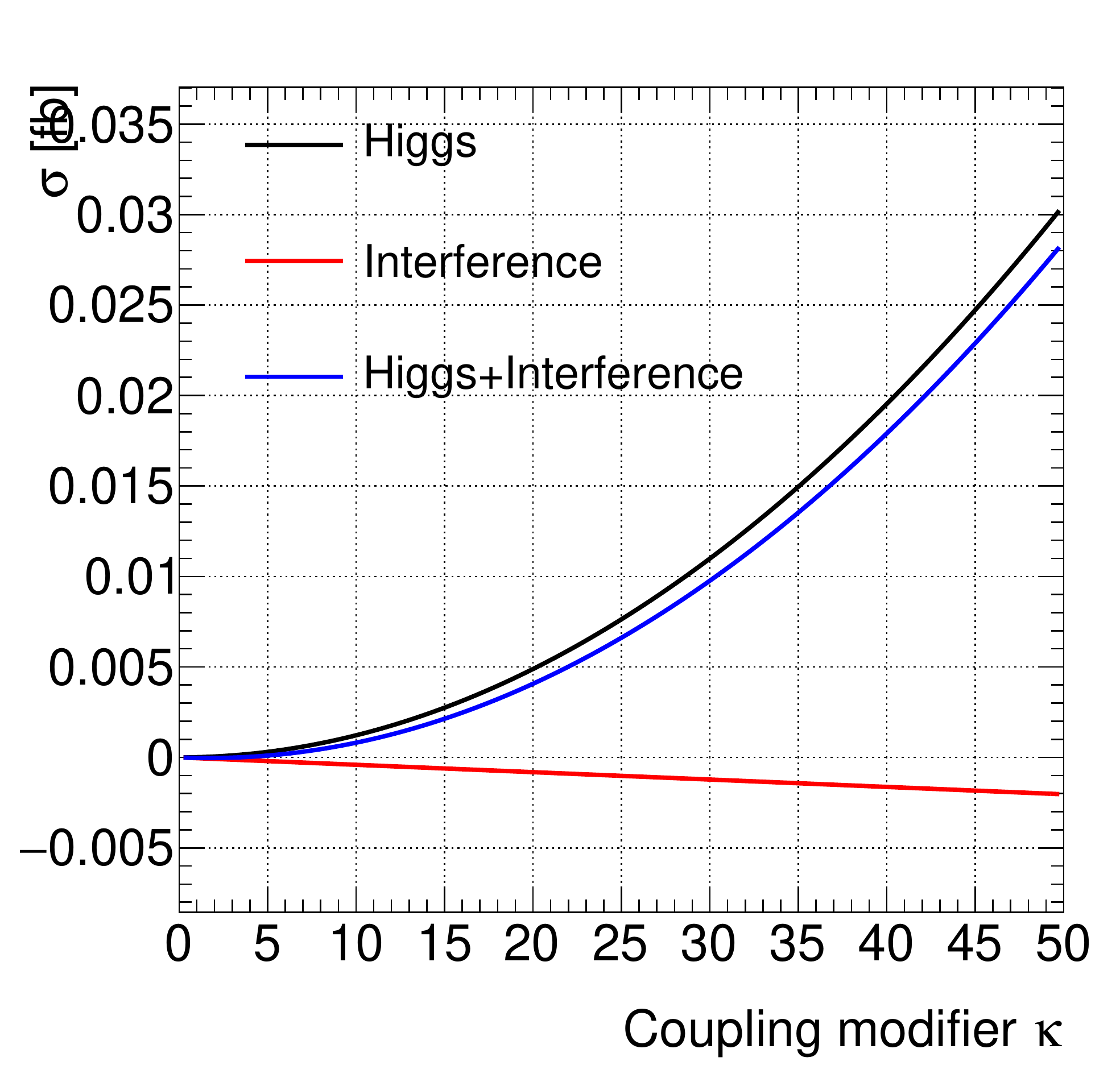}
     \caption{\label{fig:xs_ptll800}
     Total fiducial cross secrtion as a function of the coupling modifier $\kappa$ (Left) and the interference cross section (Right) for the region $\pt(ll)>800$~GeV. On the left plot, the green solid curve and red dashed curve show the fit results with or without including the interference term.
     }
 \end{figure}

We will look at the interference effect from two aspects: the total fiducial cross sections and various differential distributions. Figures~\ref{fig:xs_ptll180}~\ref{fig:xs_ptll180to400}~\ref{fig:xs_ptll400to800}~\ref{fig:xs_ptll800} show the cross sections as functions of the the coupling modifier $\kappa$ in different $\pt(ll)$ regions as well as the fitting results with or without including the interference term. The fits give the chi-square difference $\Delta \chi^2 = 21.03$ using all
available MC samples, which clearly indicates the existence of a destructive interference.  

\begin{figure}[htbp]
     \centering
     \includegraphics[width=0.32\textwidth]{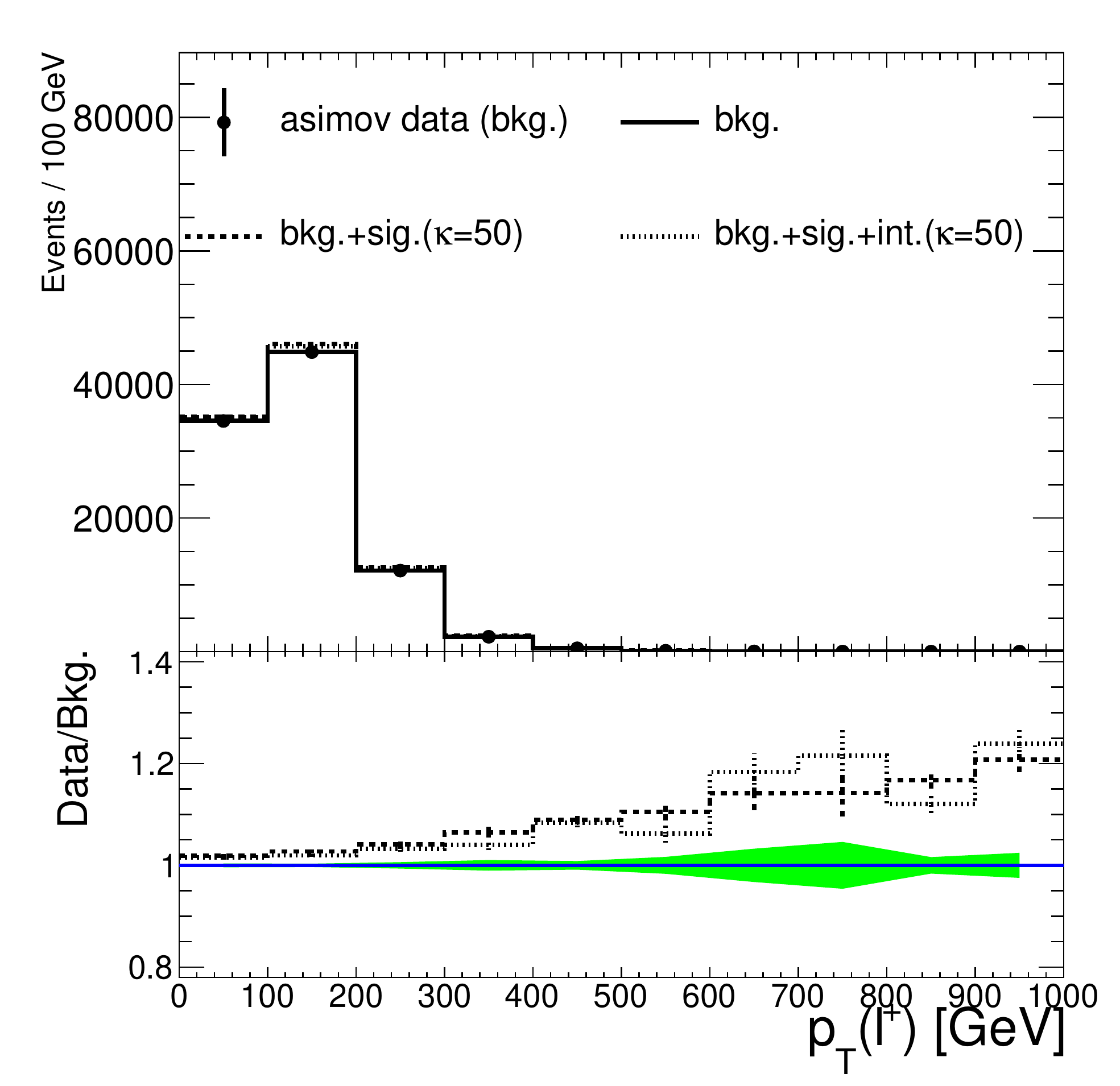}
     \includegraphics[width=0.32\textwidth]{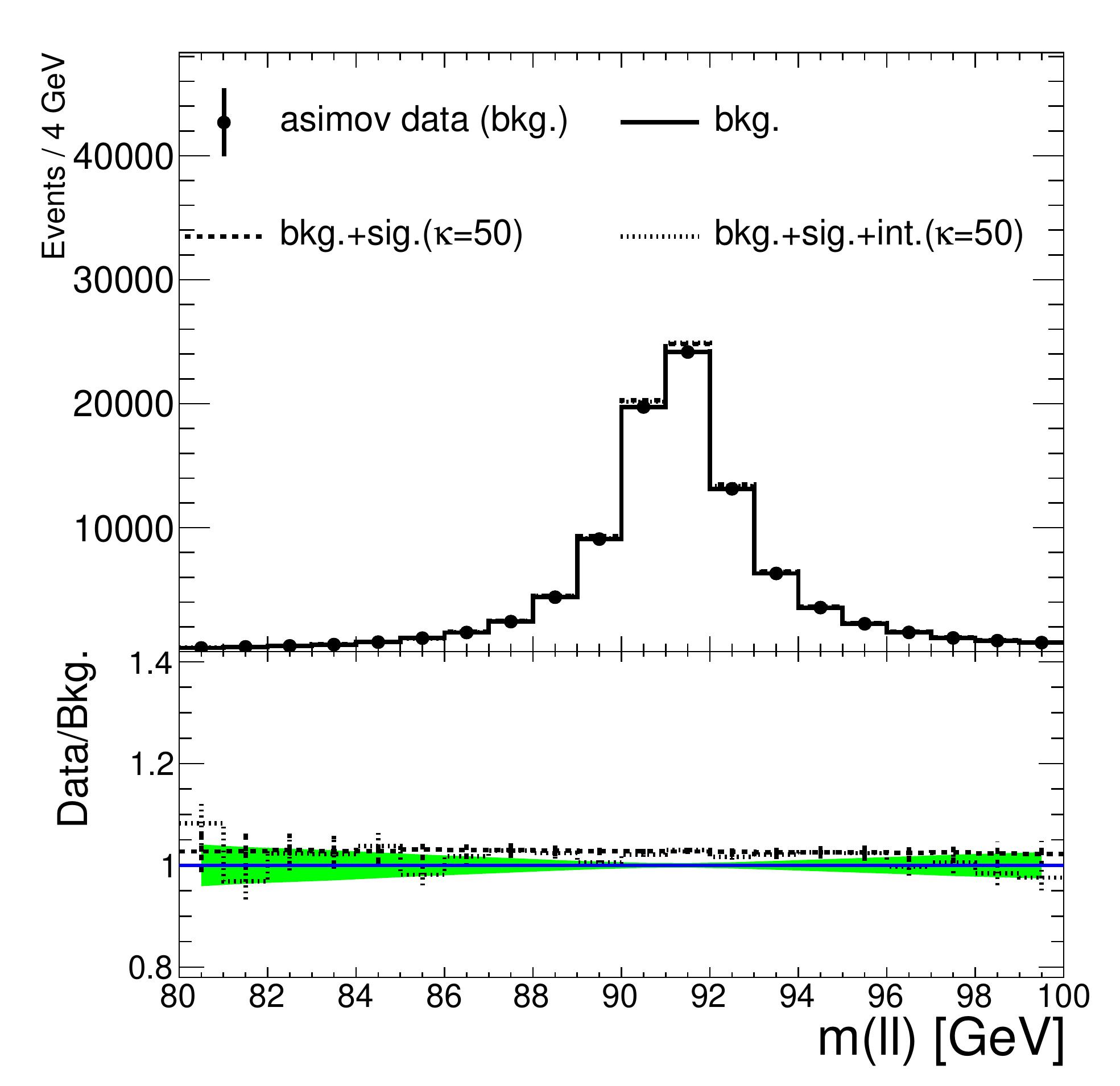}
     \includegraphics[width=0.32\textwidth]{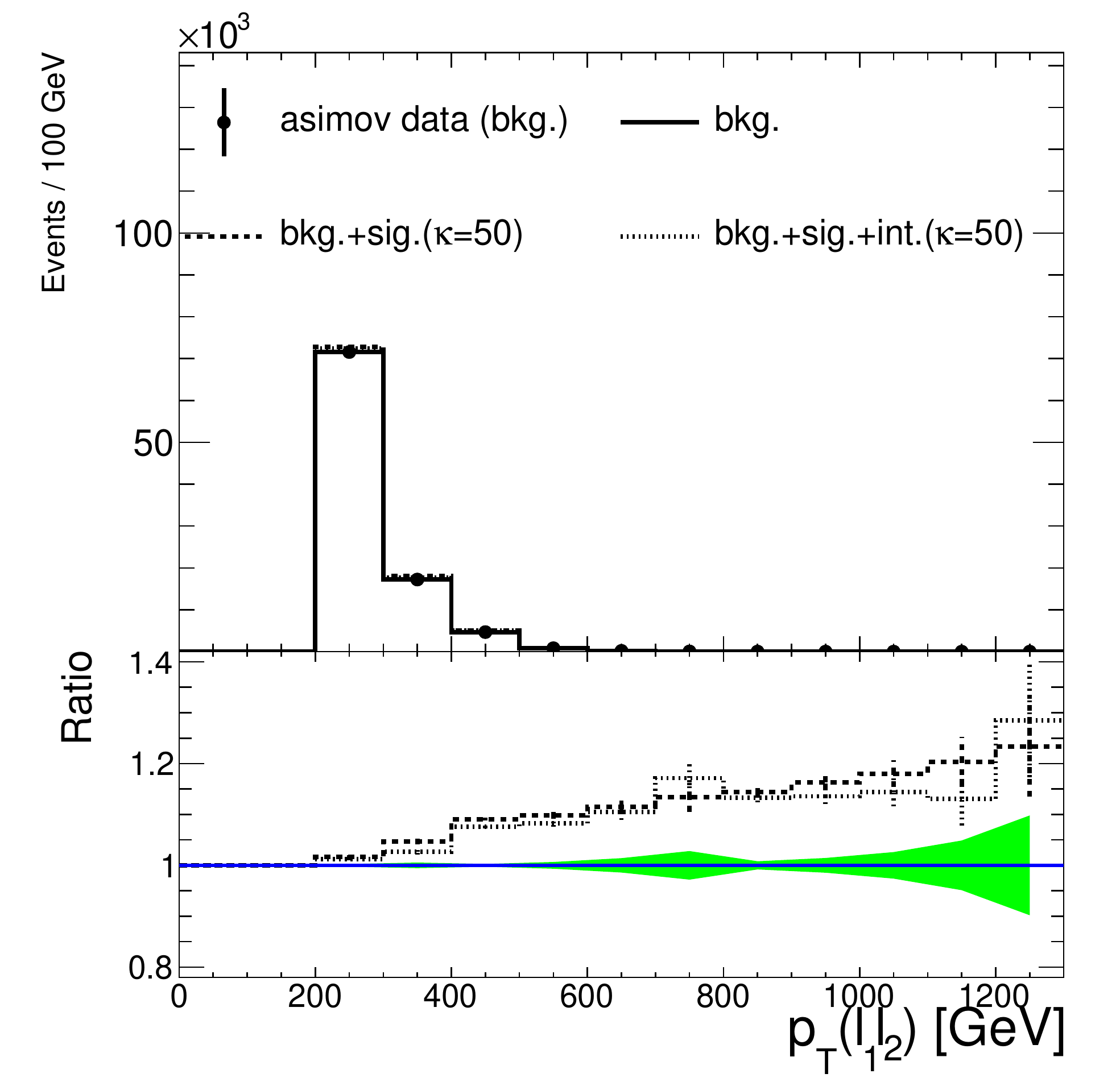}\\
     \includegraphics[width=0.32\textwidth]{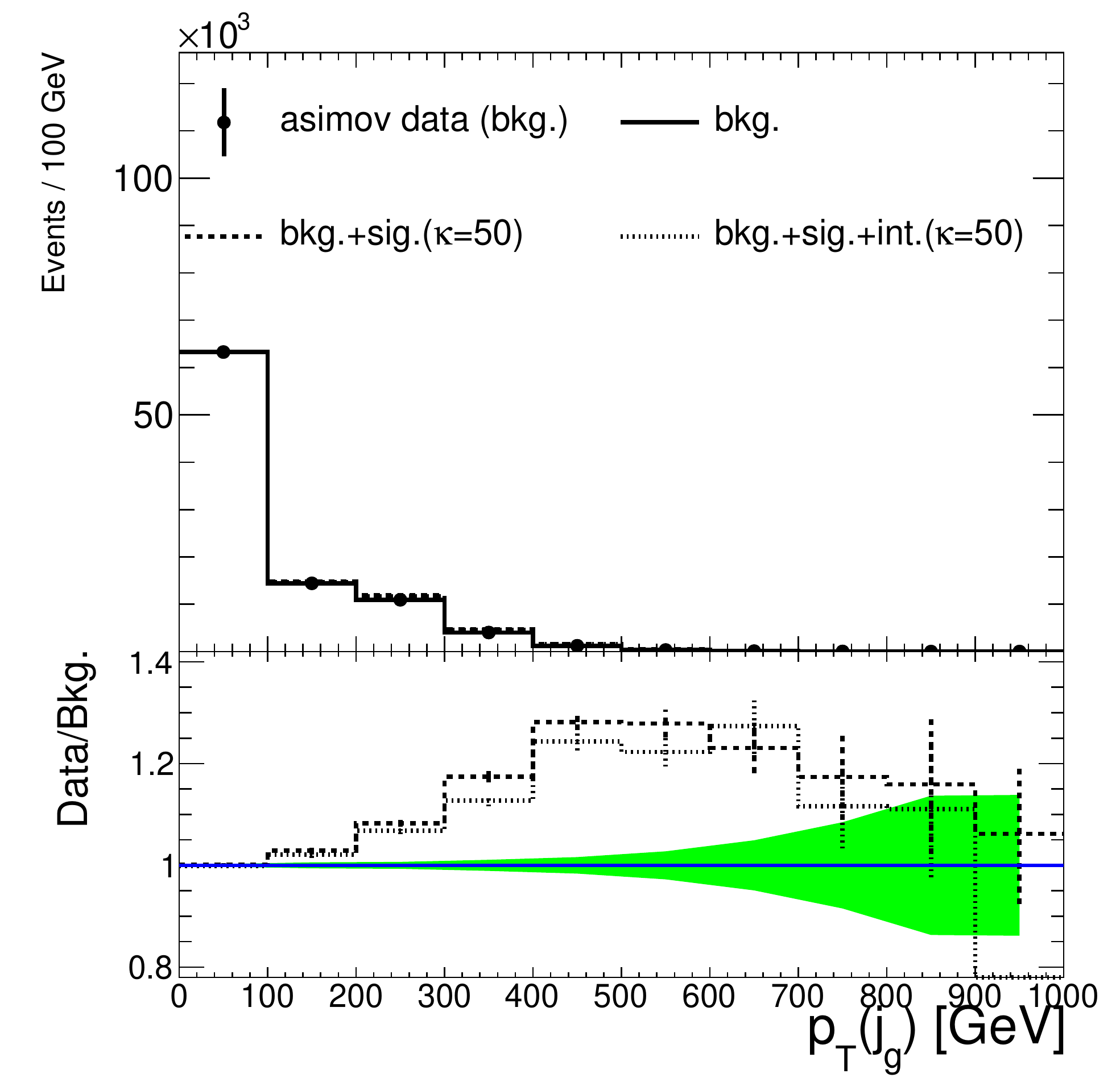}
     \includegraphics[width=0.32\textwidth]{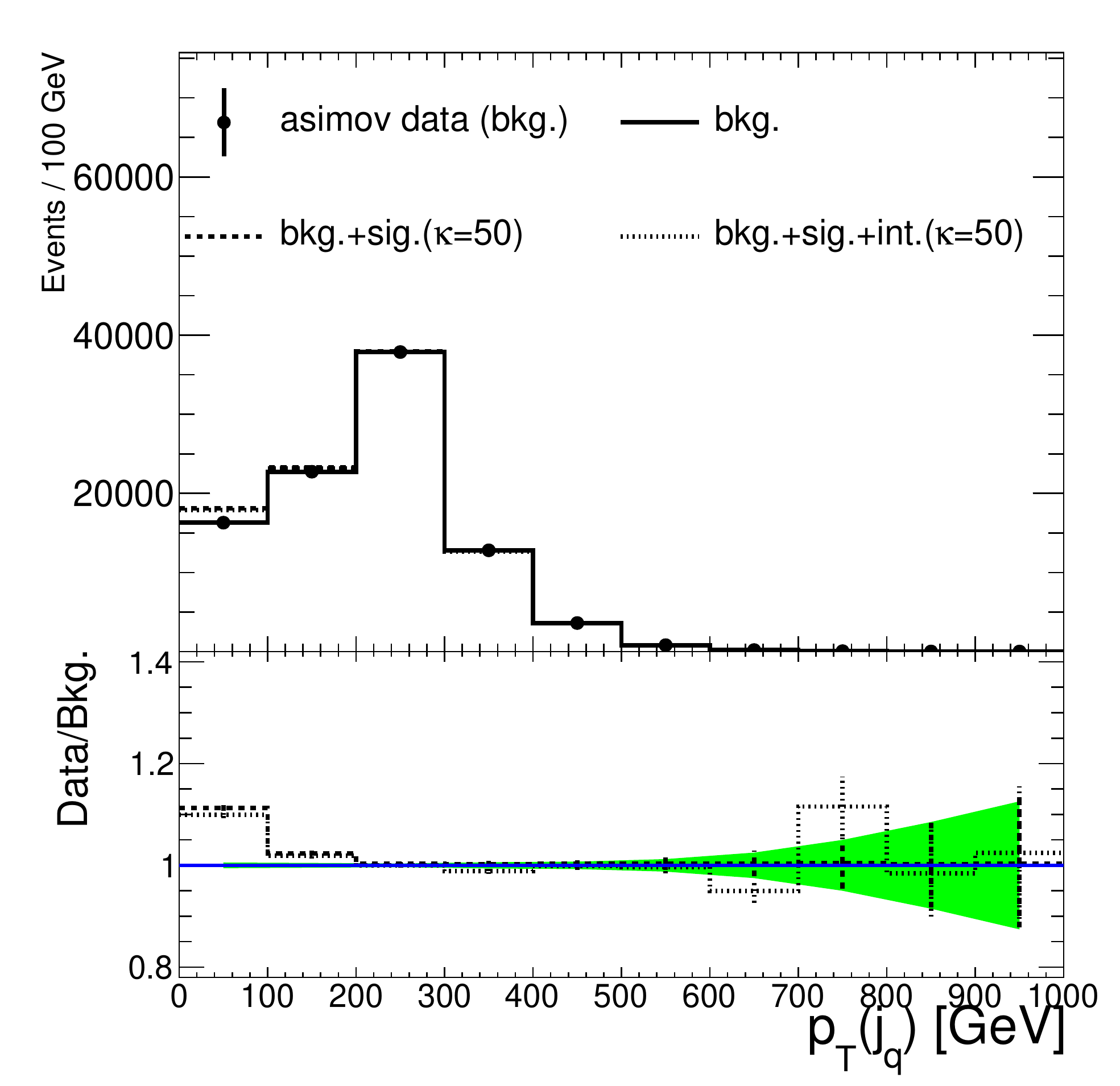}
     \includegraphics[width=0.32\textwidth]{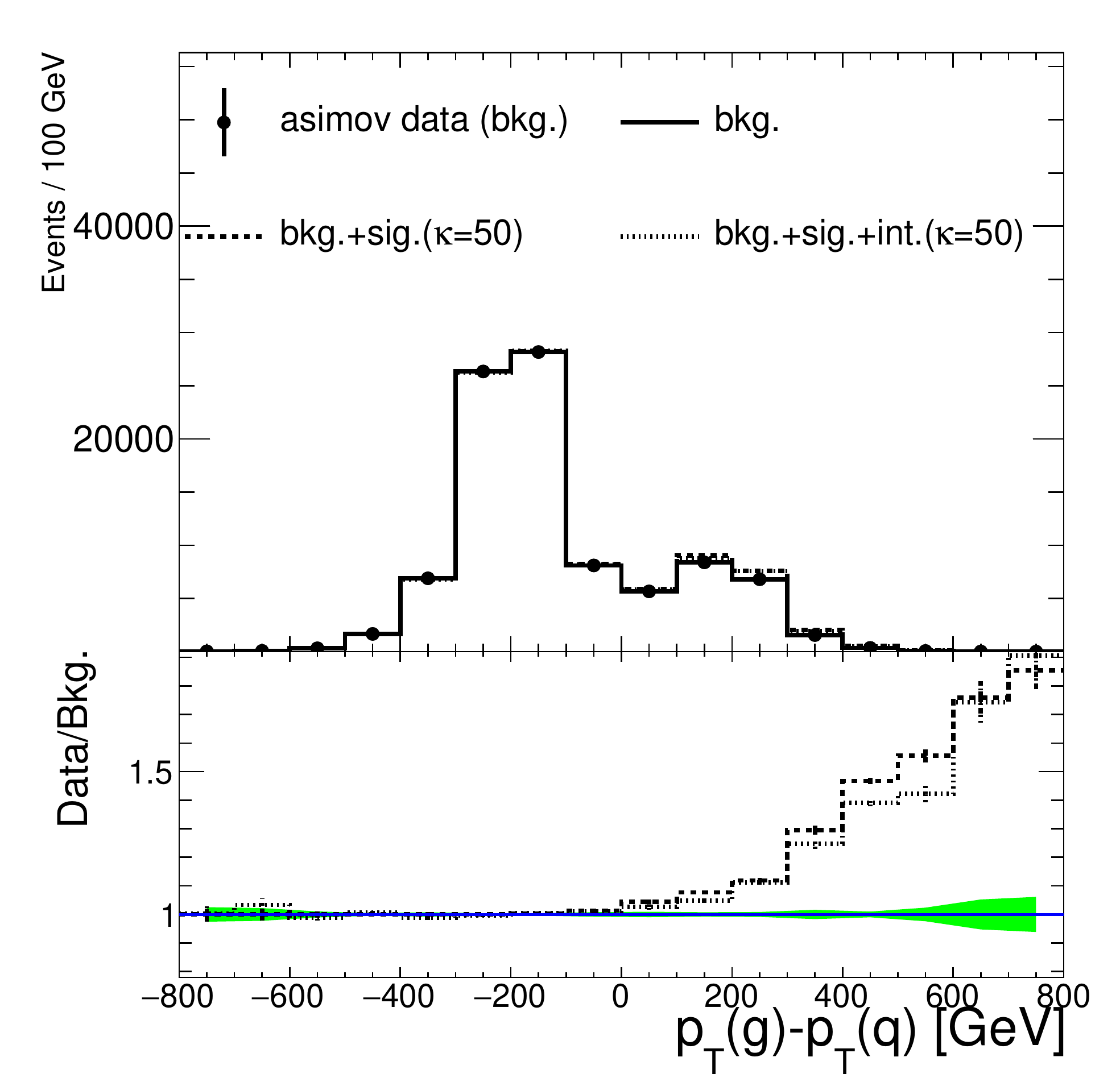}\\
     \includegraphics[width=0.32\textwidth]{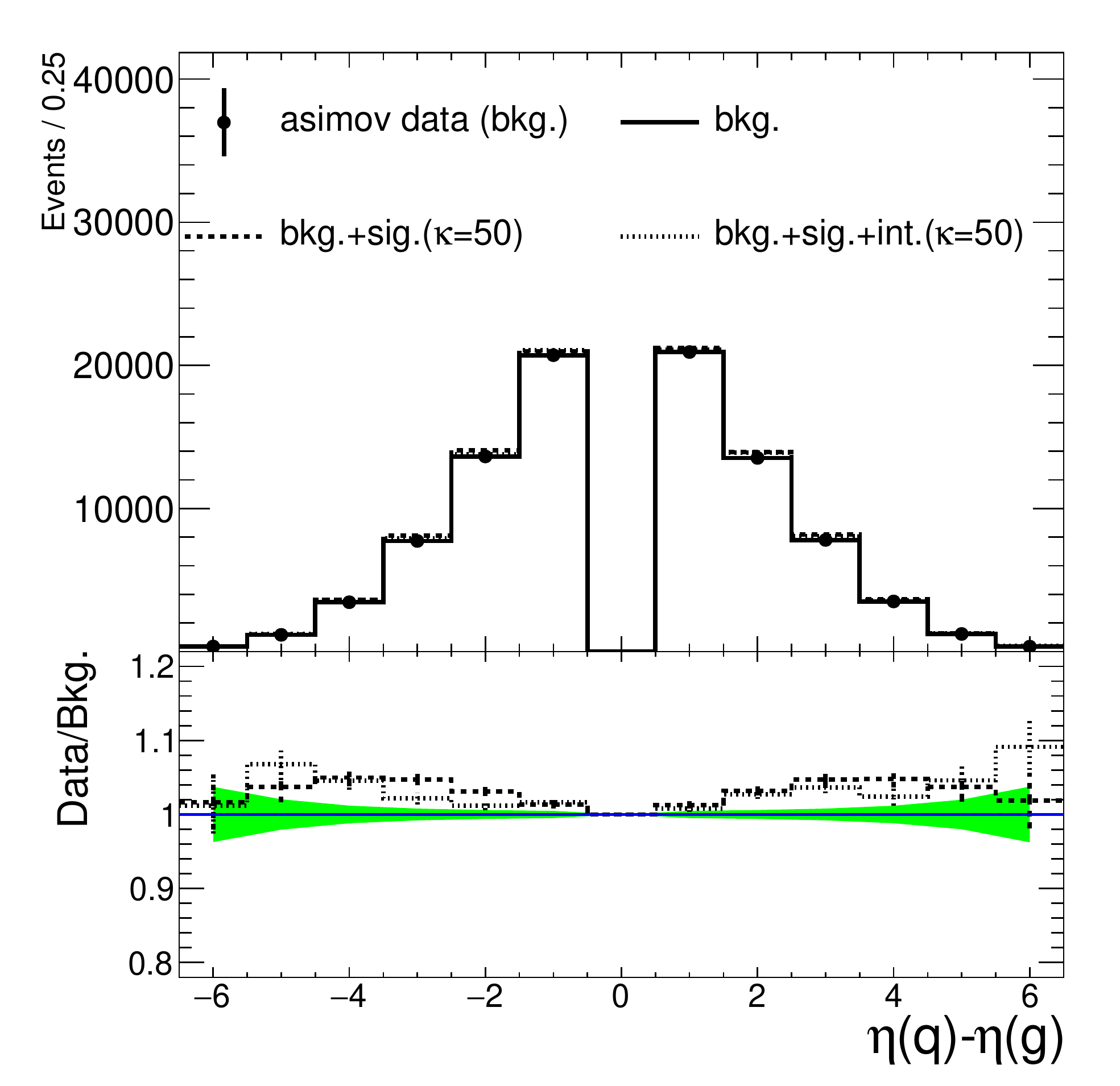}
     \includegraphics[width=0.32\textwidth]{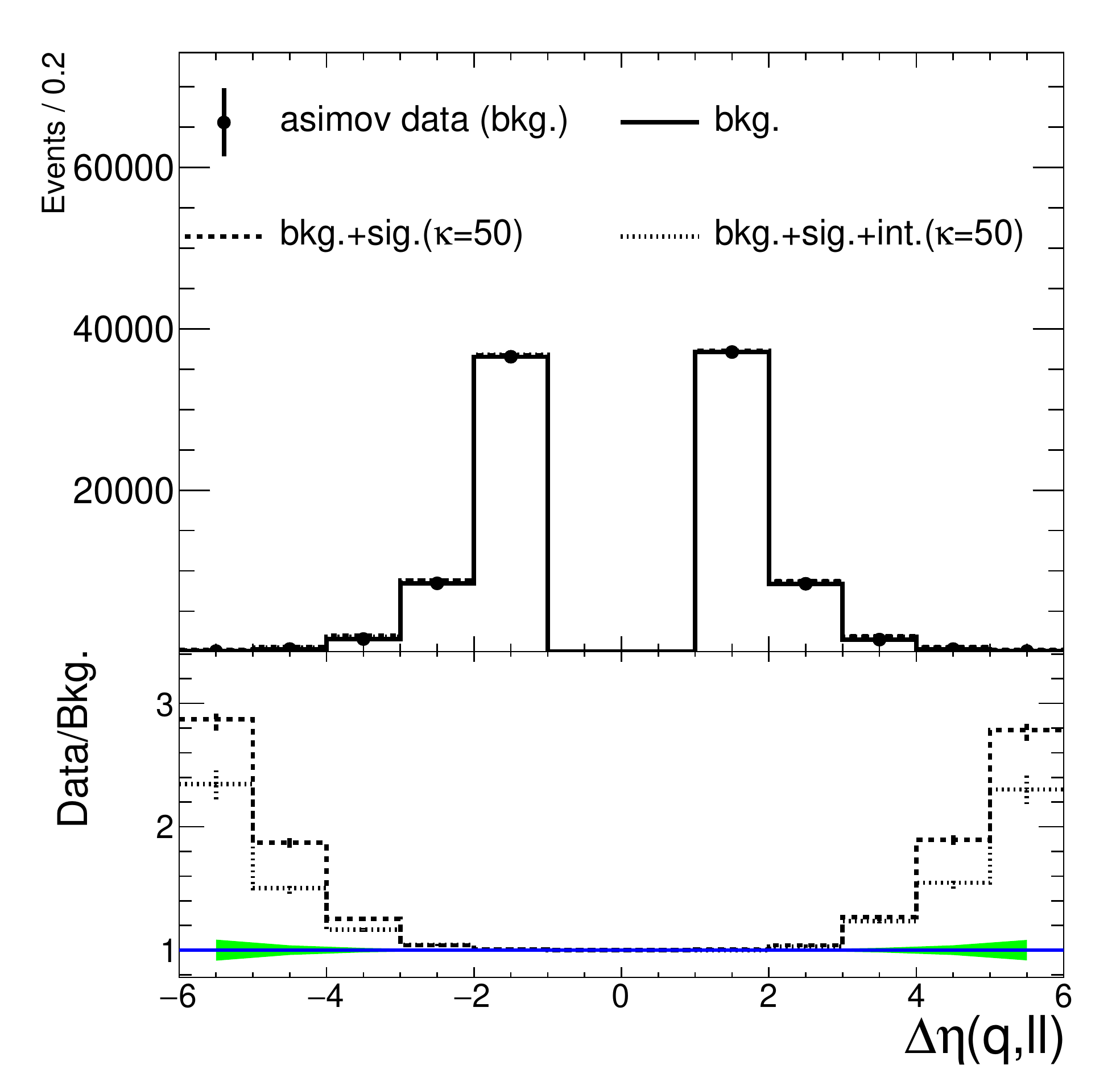}
     \includegraphics[width=0.32\textwidth]{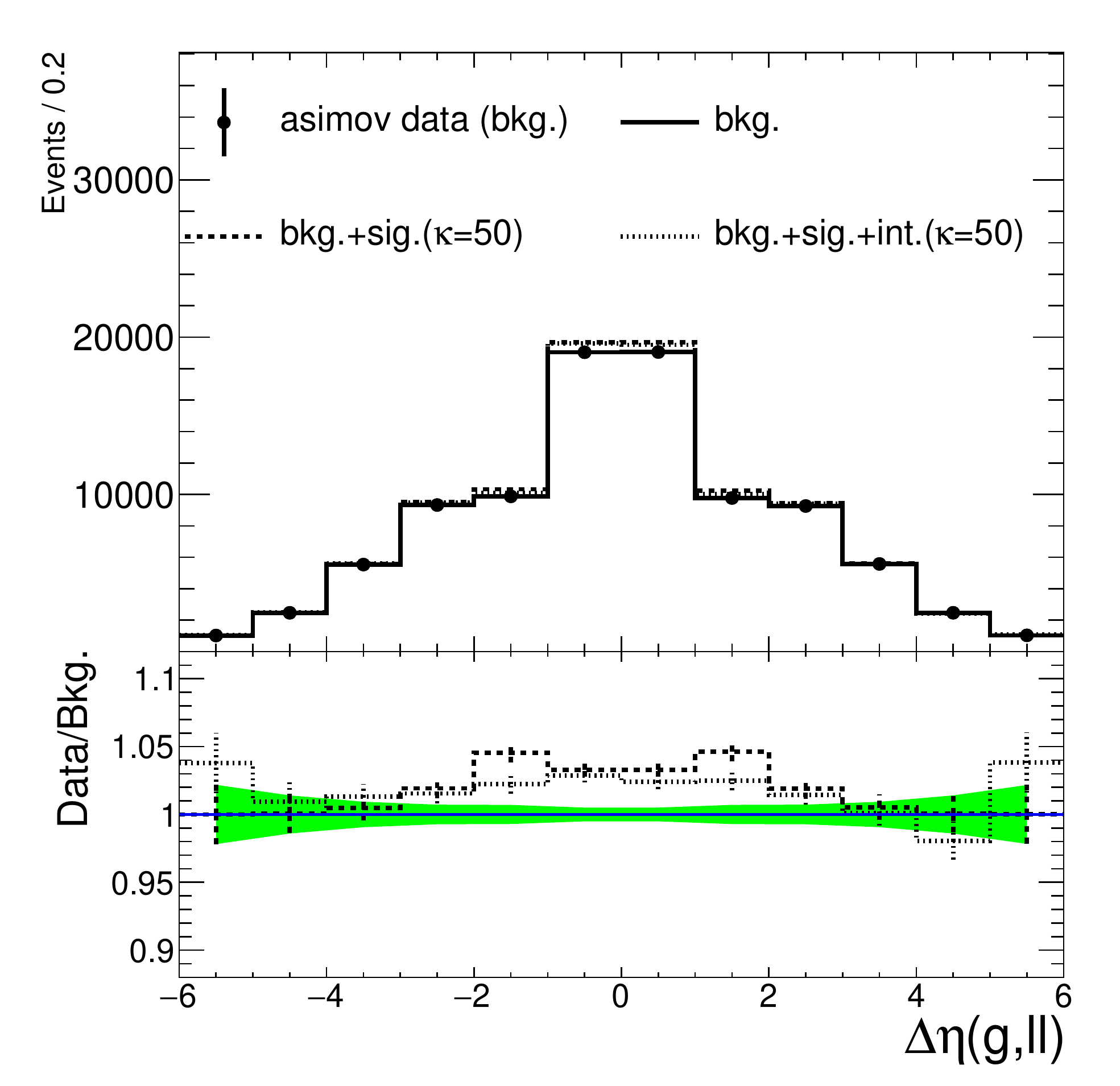}\\
     \caption{\label{fig:int_diff}
     Differential distributions of various kinematic variables. The green band in the lower pad represents the Poisson uncertainty.
     }
 \end{figure}

 To illustrate the interference effect on various kinematic variables, we take $\kappa=50$ as example and plot the differential distributions at generator level with or without including the interference. To match the event selections we will describe in next section, the cuts $\pt(l)>40$~GeV, $\pt(ll)>200$~GeV, $|m(ll)-90|<10$~GeV, $|\eta(q)-\eta(ll)|>1$ and $|\eta(q)-\eta(g)|>0.5$ are applied. Figure~\ref{fig:int_diff} shows 9 kinematic variables. Some of them like $\pt(g)-\pt(q)$ and
 $\pt(ll)$ are sensitive to the signal. Significant interference effect is only seen on $\eta(g)-\eta(ll)$ and $\eta(q)-\eta(ll)$. As the current dataset collected by ATLAS or CMS is not enough for us to really probe the minor interference effect, we will neglect the interference effect in the following sections. 

\section{Signal topology and event selection}
The topology of the signal process $g+q \to g+q +l^++l^-$ driven by a $t$-channel Higgs boson has 4 features: 1) the $\pt$ spectrum of the leptons is hard; 2) the $\pt$ spectrum of the gluon jet is hard while the $\pt$ spectrum of the quark is soft; 3) the rapidity between the gluon jet and the quark jet is large; 4) the rapidity distance between the di-lepton system and the quark jet is large. The dominant background has opposite features: the leptons' $\pt$ spectrum is
relatively soft; the quark jet's $\pt$ spetrum is harder than the gluon; the angular distance between the quark and the gluon is
relatively closer because the gluon may be radiated from the quark; the angular distance between the quark and the di-lepton system is also closer because the virtual photon or $Z$ boson may be radiated from the quark. 

In the following analyis, we will only study the case of $l=\mu$ and take $\kappa=50$ for illustration. Signal and background MC samples are produced using the MadGraph generator. Parton showering is performed using Pythia8 and the detector response is simulated using Delphes based on the ATLAS performance in Run~II. Events with at least two jets with $\pt>20$~GeV and two charged muons with $\pt>5$~GeV and opposite charge sign are selected.   

\begin{figure}[htbp]
     \centering
     \includegraphics[width=0.45\textwidth]{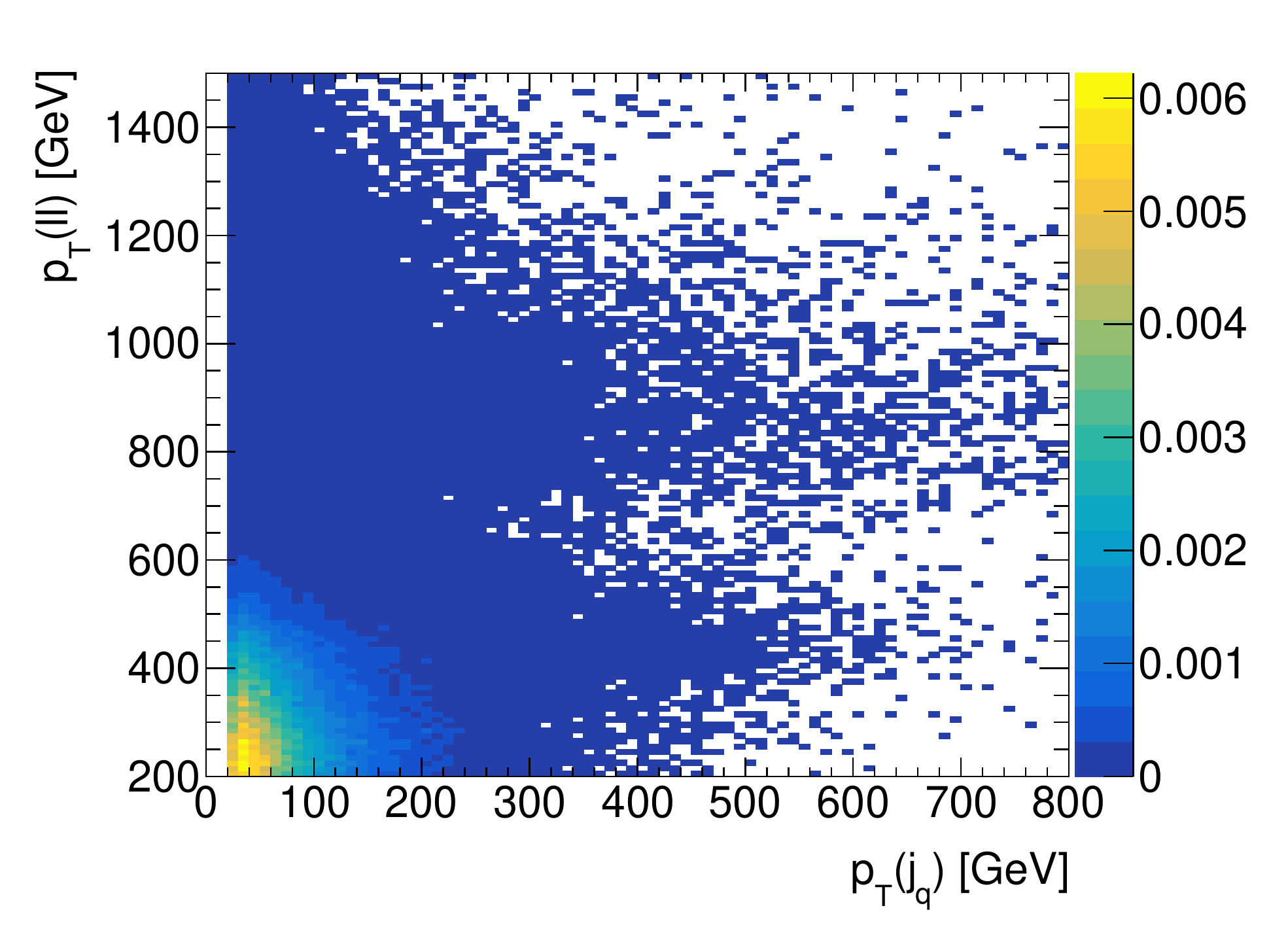}
     \includegraphics[width=0.45\textwidth]{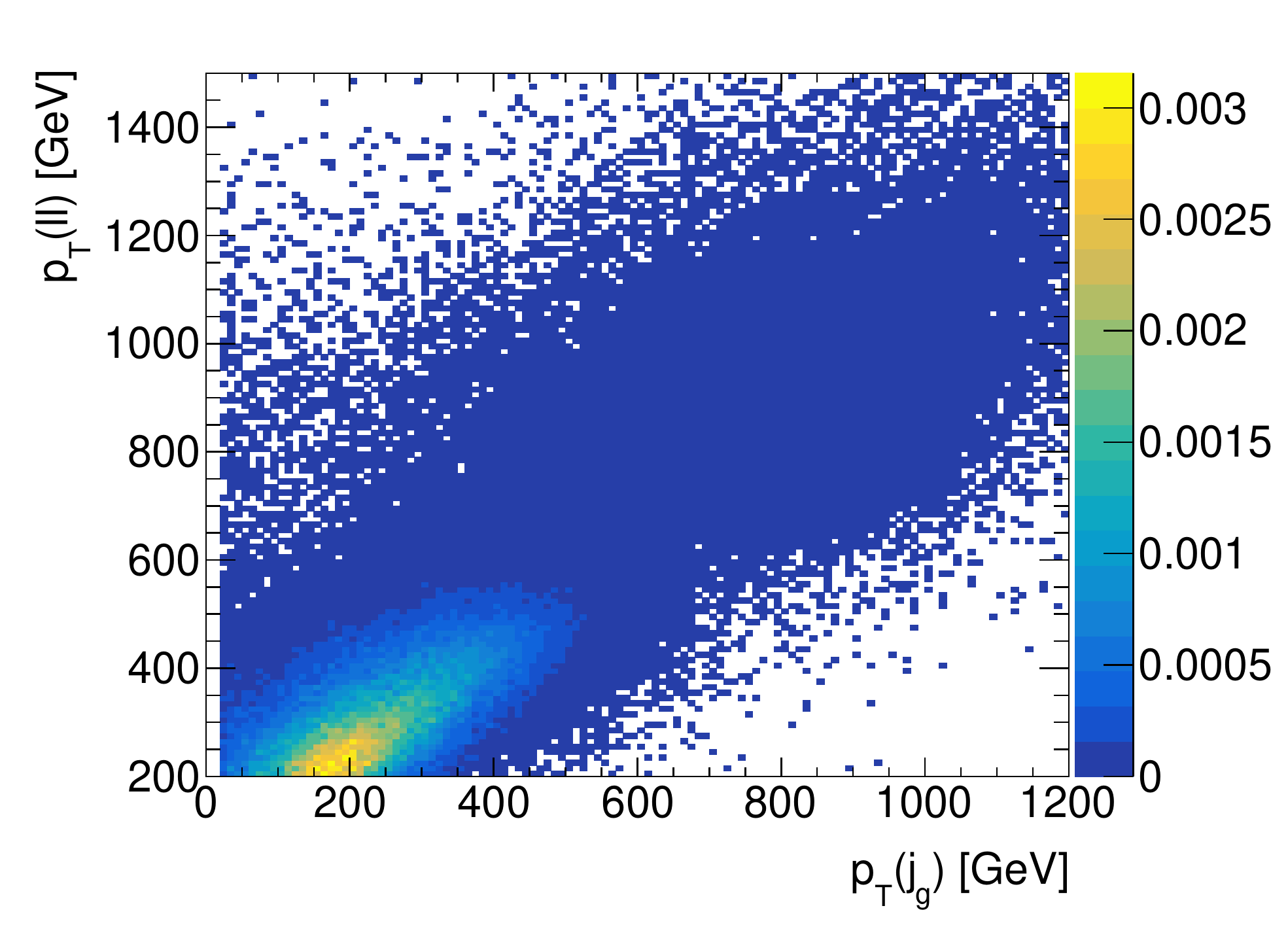}\\
     \includegraphics[width=0.45\textwidth]{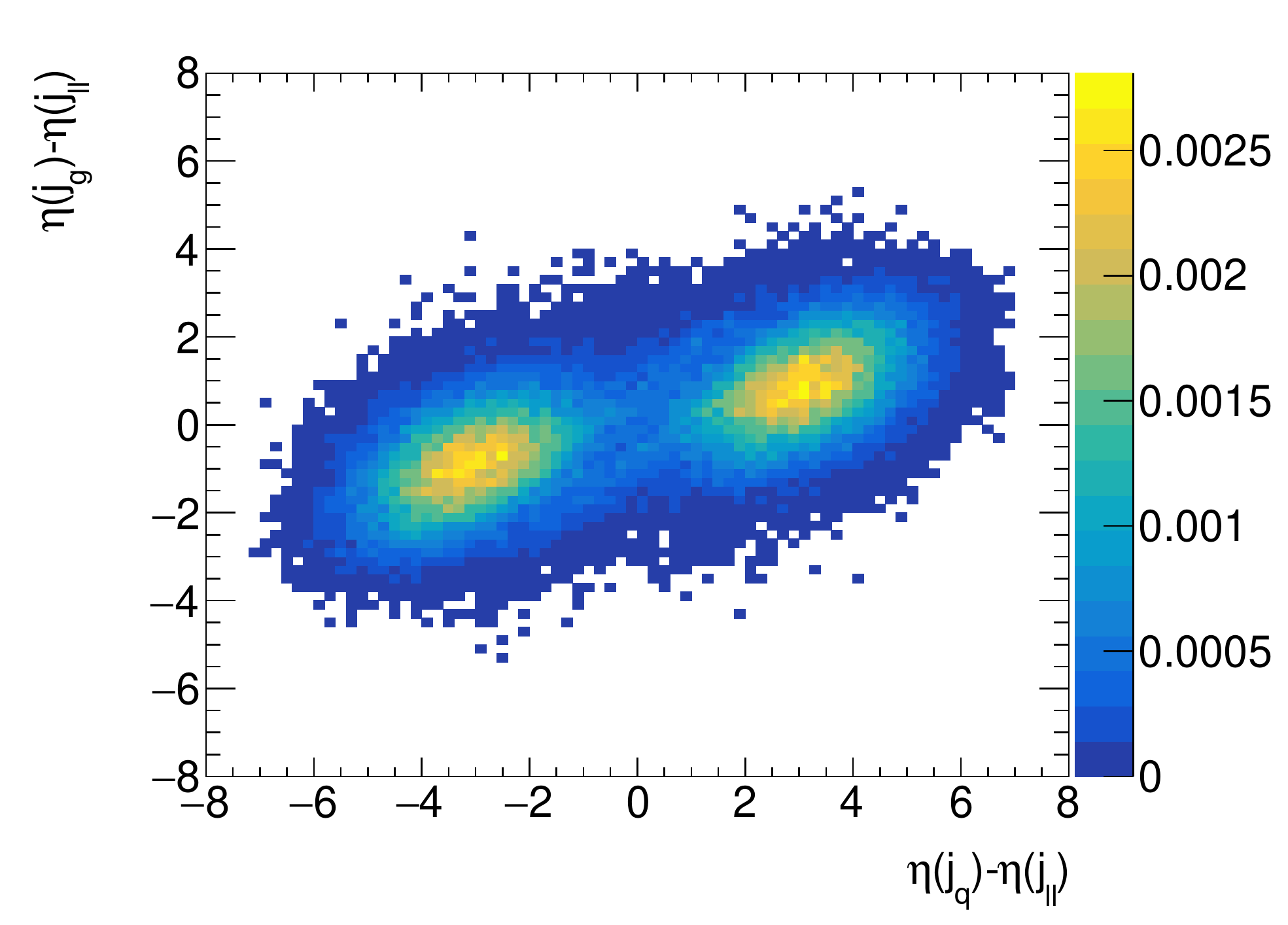}
     \includegraphics[width=0.45\textwidth]{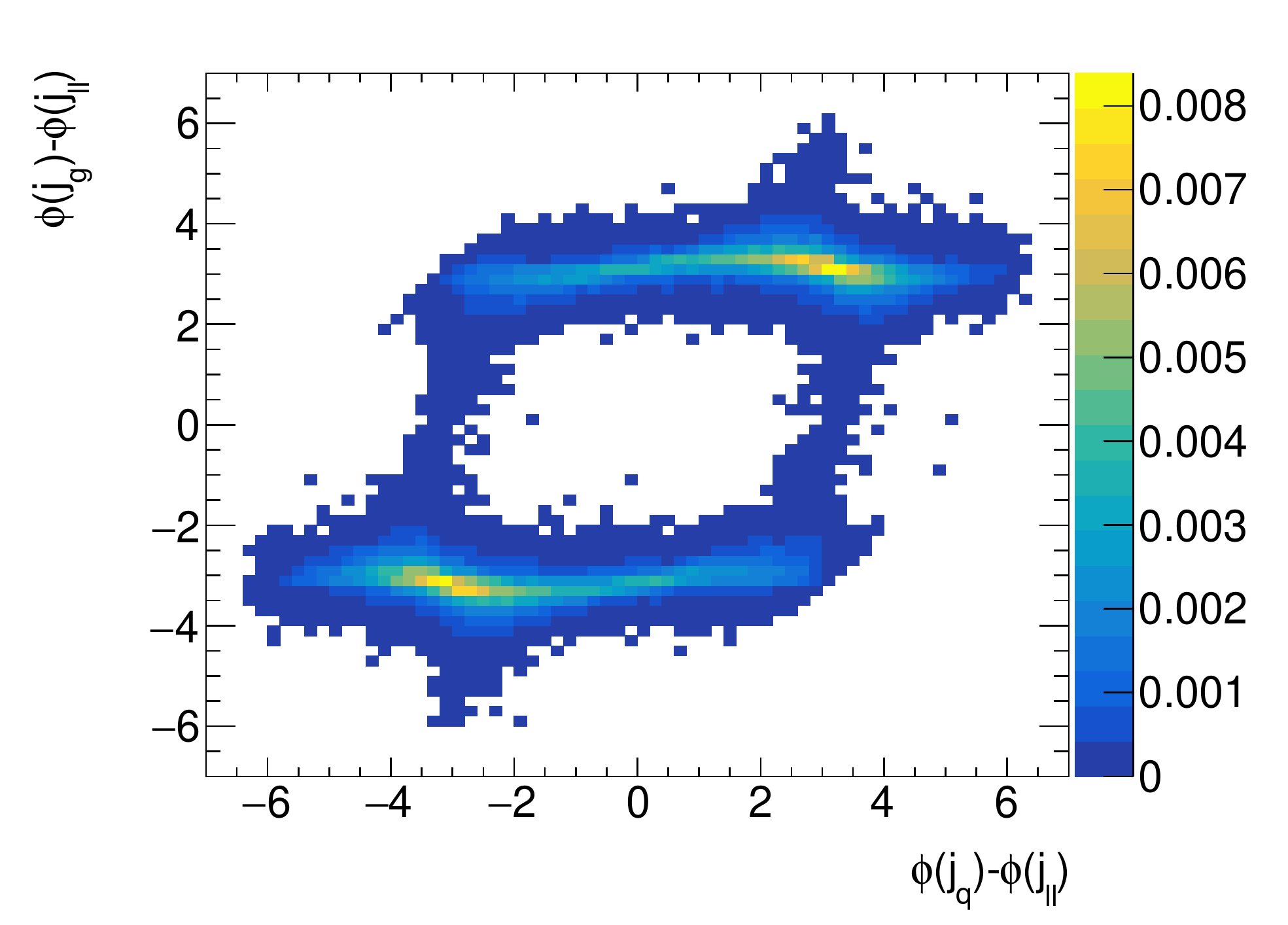}
     \caption{\label{fig:jet_match}
     The distributions of $\pt(j_q):\pt(ll)$ (Top Left), $\pt(j_g):\pt(ll)$ (Top Right), $\Delta\eta(q,ll):\Delta\eta(g,ll)$ (Bottom Left) and $\Delta\phi(q,ll):\Delta\phi(g,ll)$ (Bottom Right) in the signal sample. 
     }
 \end{figure}

\begin{figure}[htbp]
     \centering
     \includegraphics[width=0.45\textwidth]{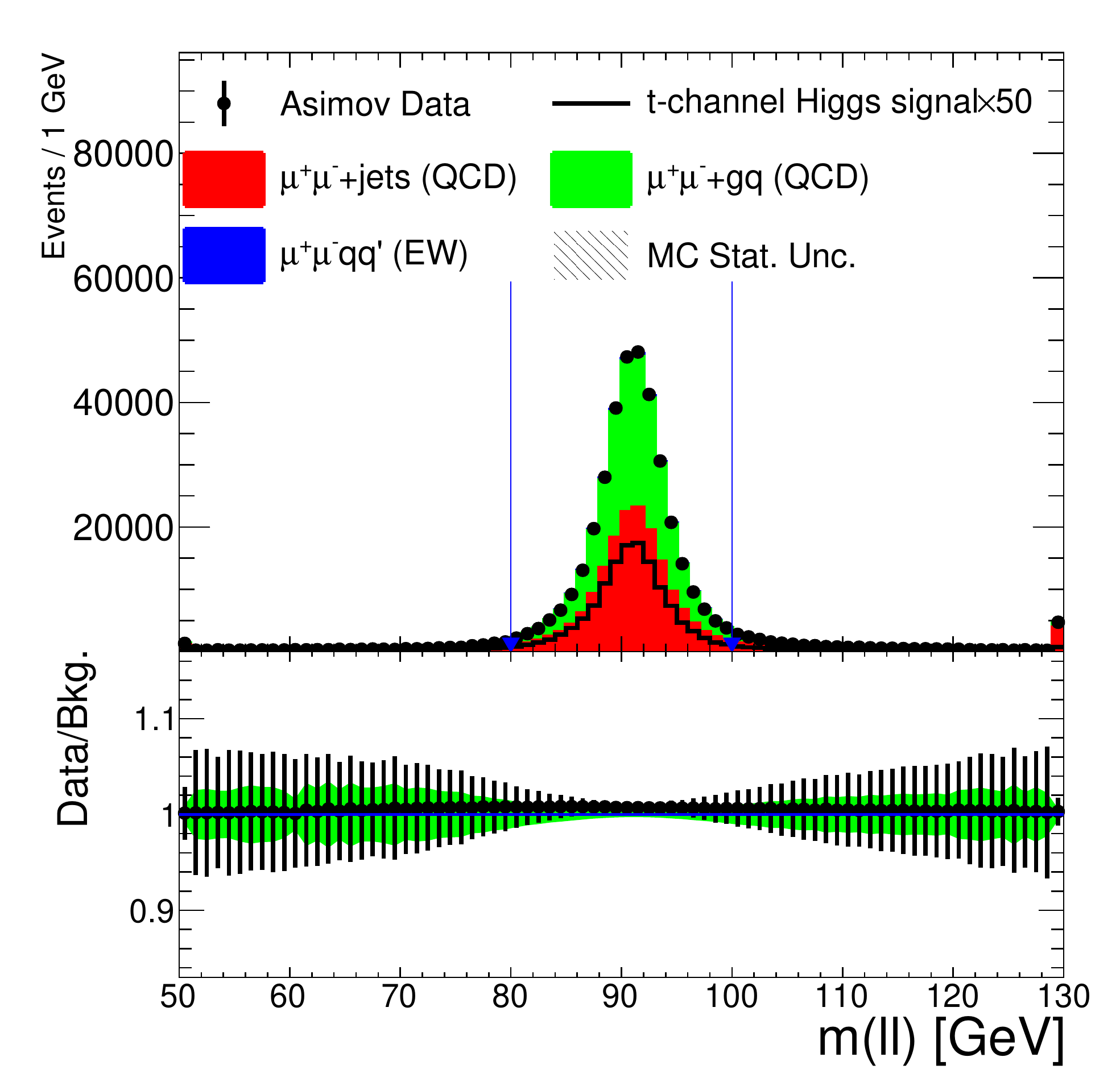}
     \includegraphics[width=0.45\textwidth]{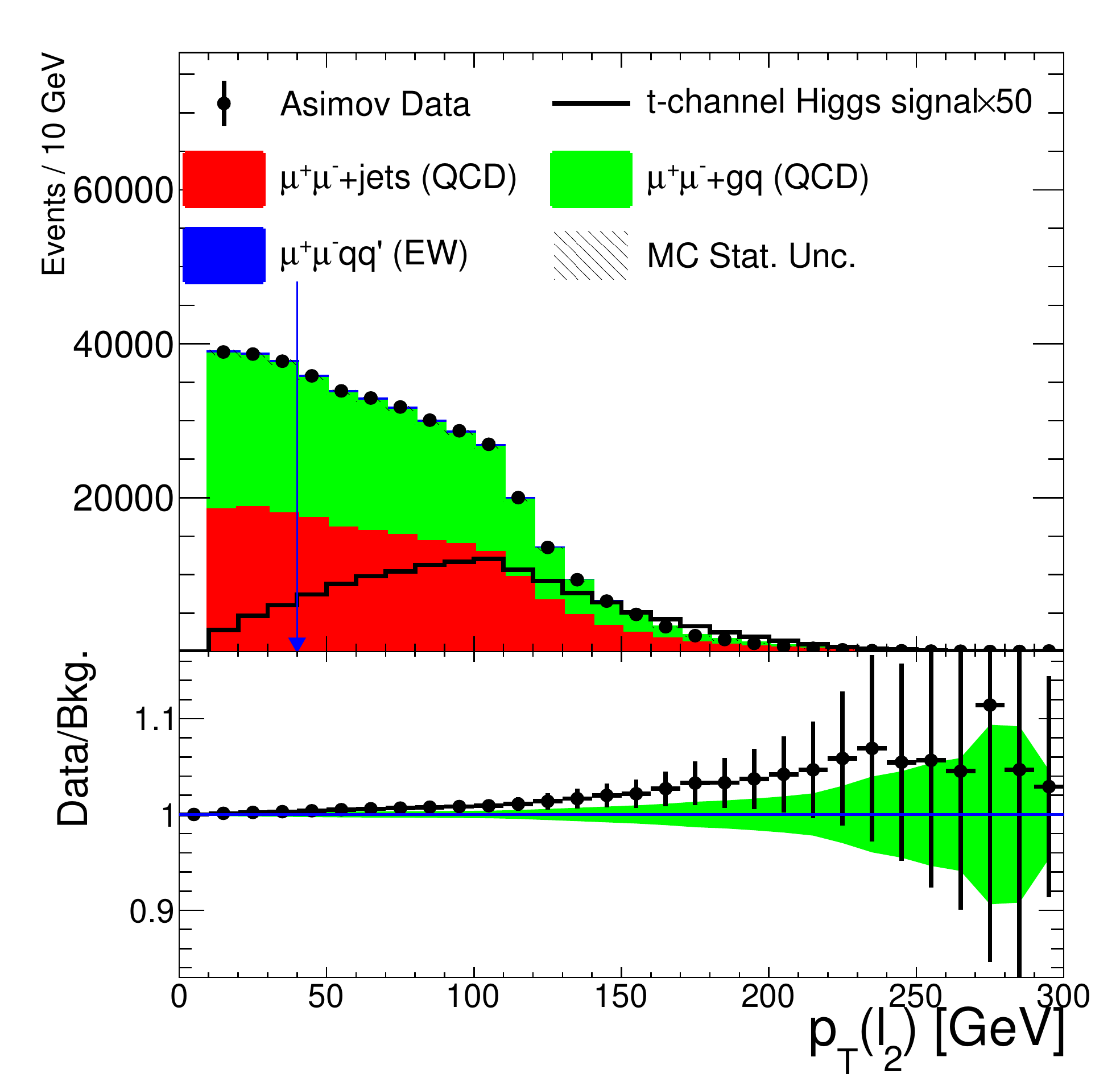}\\
     \includegraphics[width=0.45\textwidth]{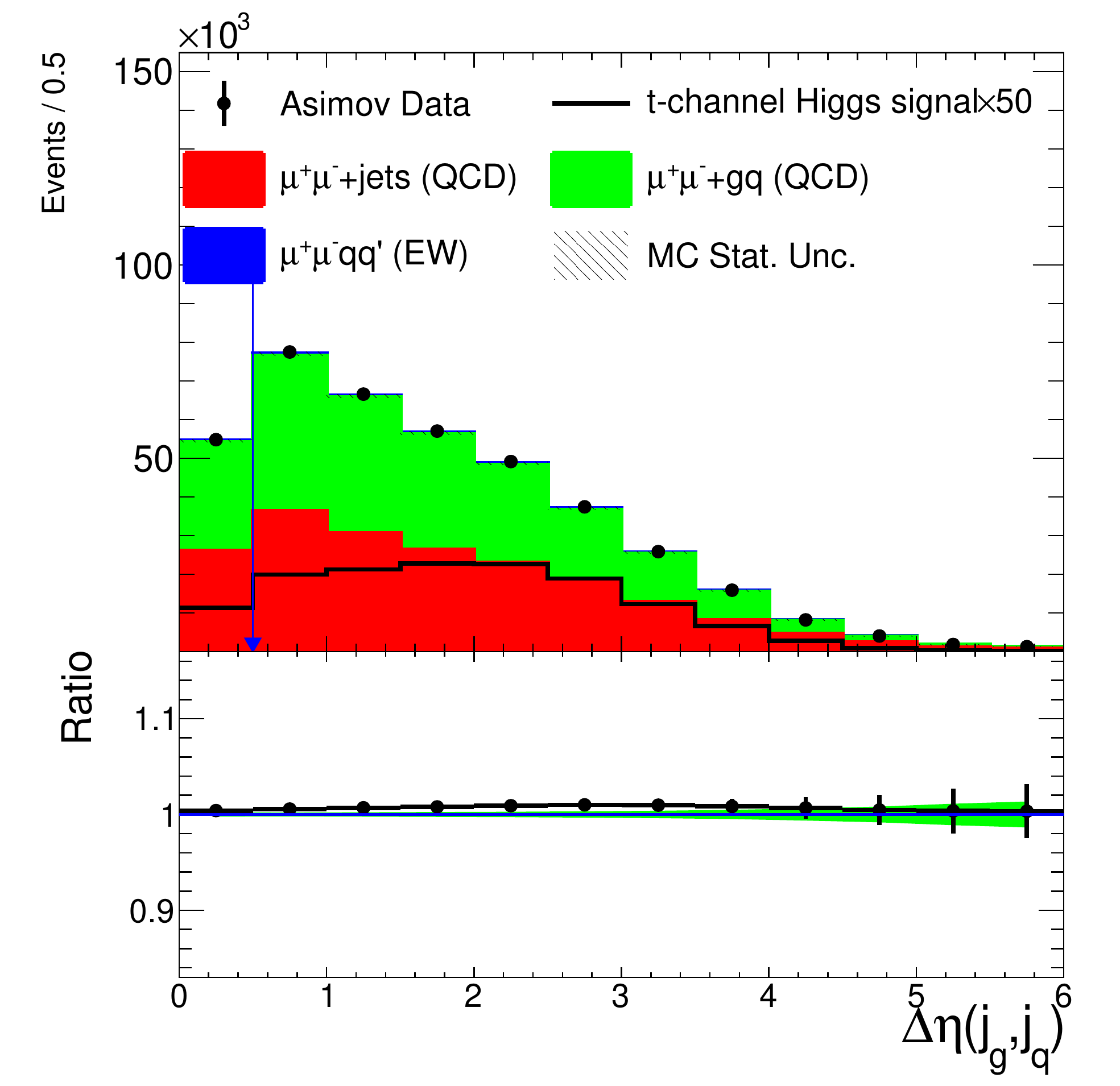}
     \includegraphics[width=0.45\textwidth]{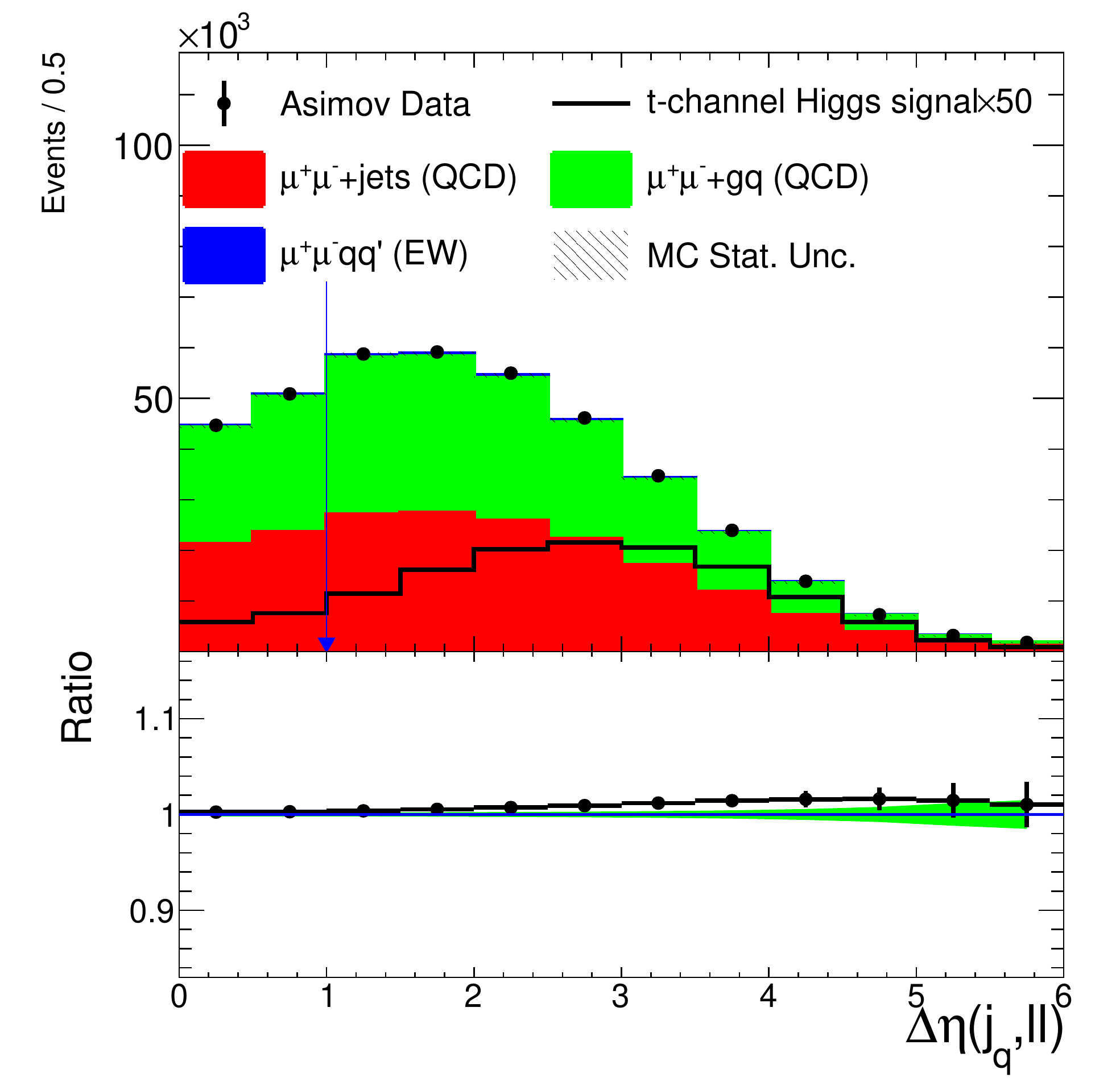}
     \caption{\label{fig:cuts}
     The distributions of $m(ll)$ (Top Left), $\pt(l_2)$ (Top Right), $|\eta(j_g)-\eta(j_q)|$ (Bottom Left) and $|\eta(j_q)-\eta(ll)|$ (Bottom Right) and the corresponding cuts. The asimov data uses $\kappa=50$. 
     }
 \end{figure}

As emphasized in last section, the signal and background have different features on the gluon jet and the quark jet. Hence it is important to identify the jet flavour as correctly as possible. A ``topology-likelihood'' technique is then developed to find the gluon jet and quark jet candidates which best match the signal topology. We choose four pairs of kinematic variables: $\pt(j_q):\pt(ll)$, $\pt(j_g):\pt(ll)$, $\Delta\eta(j_q,ll):\Delta\eta(j_g,ll)$ and
$\Delta\phi(j_q,ll):\Delta\phi(j_g,ll)$. For each pair, a 2-Dimension probability distribution is built. The best jet candidates are selected to maximize the product of the probability on each pair of kinematic variables. The probability distributions are shown in Fig.~\ref{fig:jet_match}. After the jet matching, we apply the following cuts: $\pt(l)>40$~GeV, $|\eta(g)-\eta(q)|>0.5$ and $|\eta(q)-\eta(ll)|>1$. We further require $|m(ll)-90|<10$~GeV to guarantee that the process is probing the coupling $y_Z$. The effectiveness of these cuts is shown in Fig.~\ref{fig:cuts}. 

\section{Mutlti-Variate Analysis and Sensitivity}
\begin{figure}[htbp]
     \centering
     \includegraphics[width=0.32\textwidth]{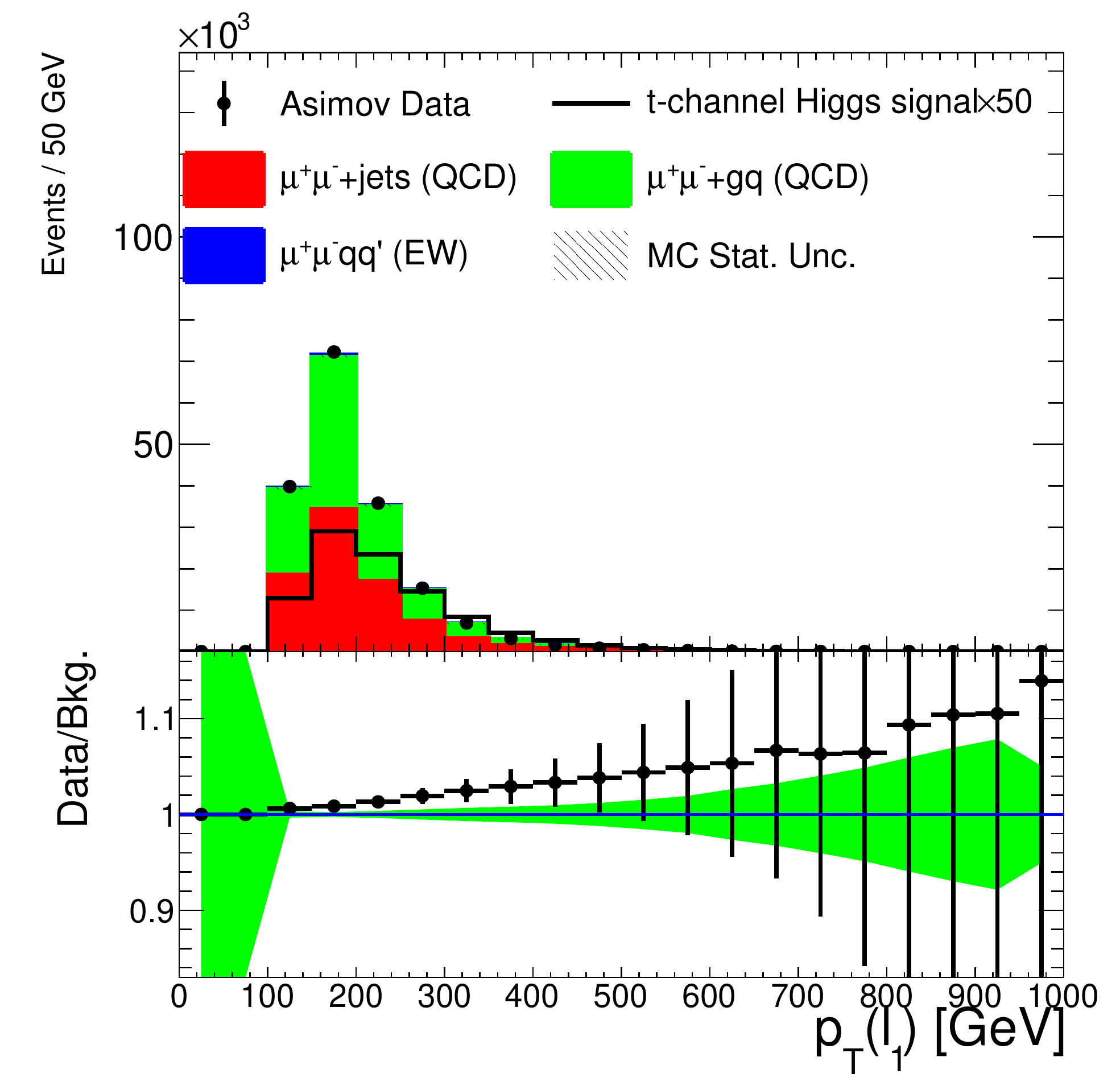}
     \includegraphics[width=0.32\textwidth]{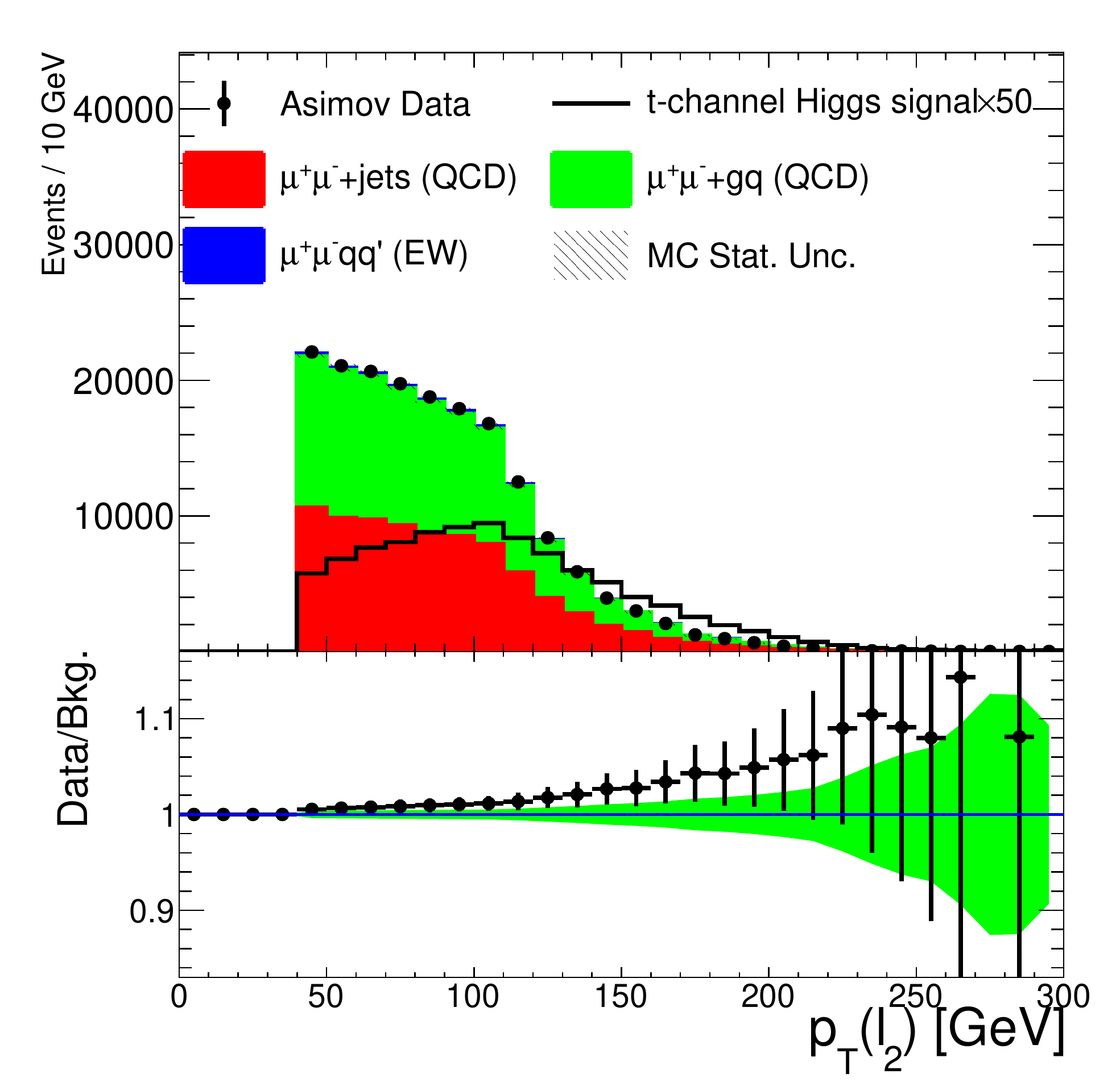}
     \includegraphics[width=0.32\textwidth]{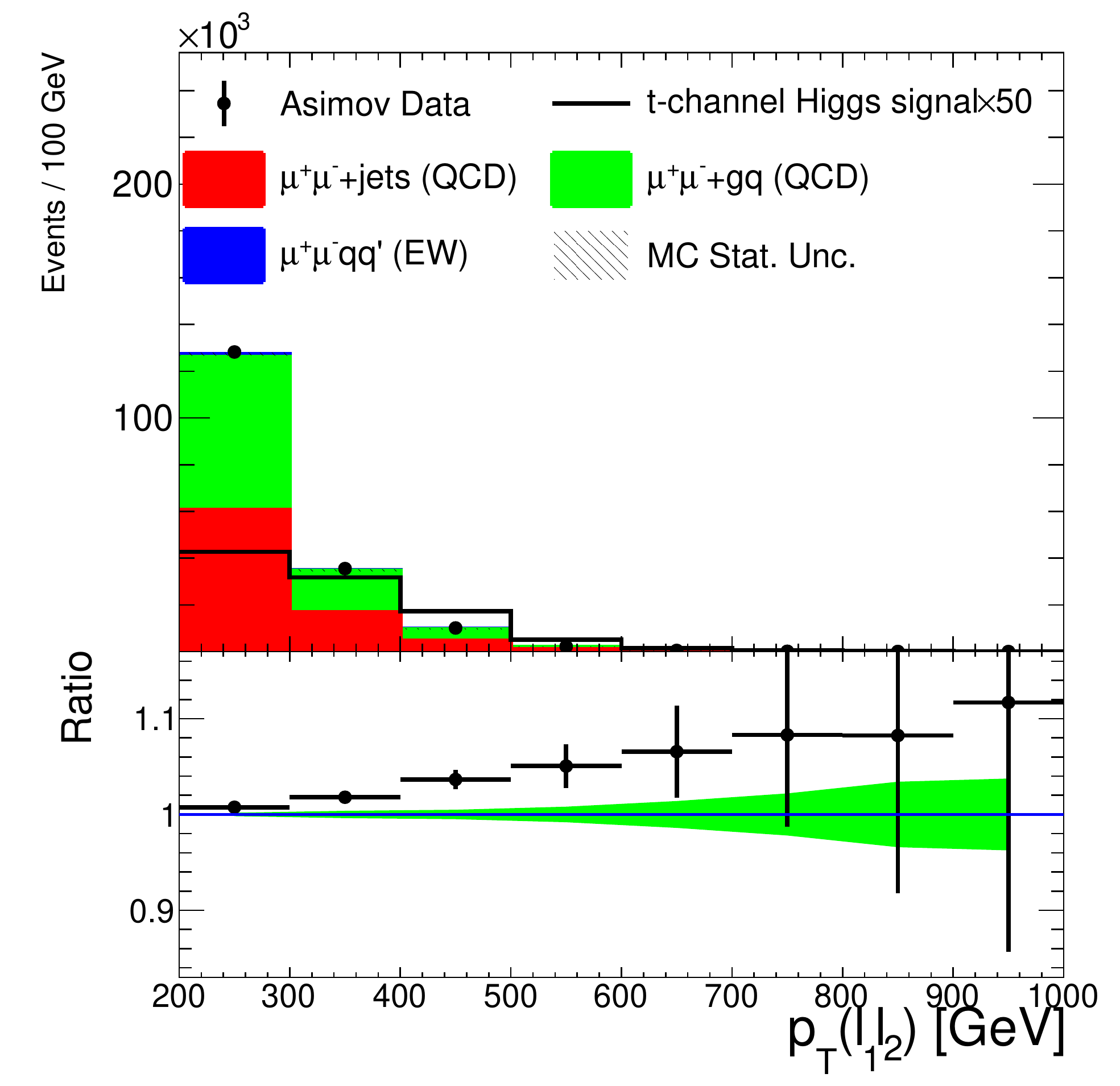}\\
     \includegraphics[width=0.32\textwidth]{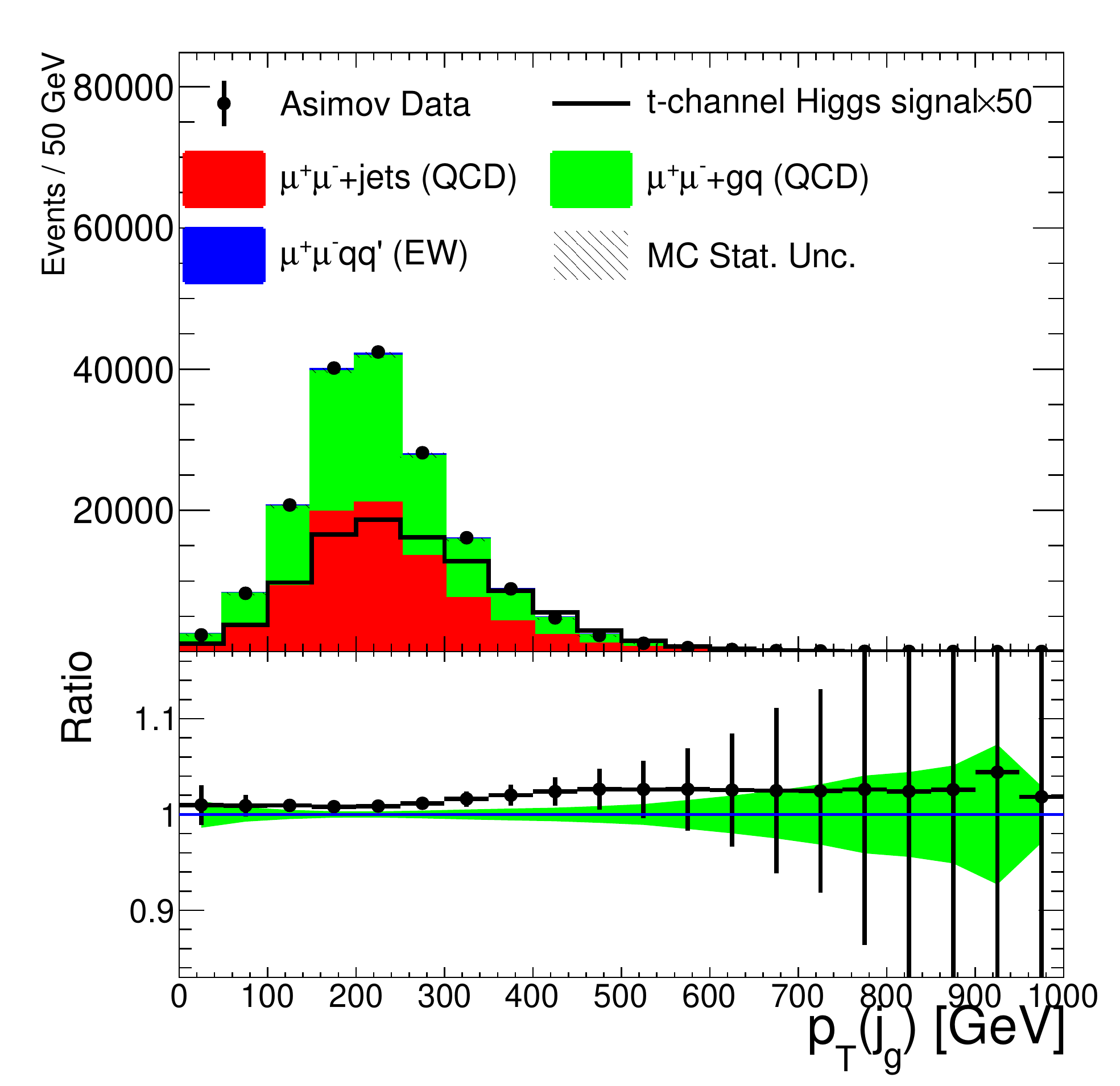}
     \includegraphics[width=0.32\textwidth]{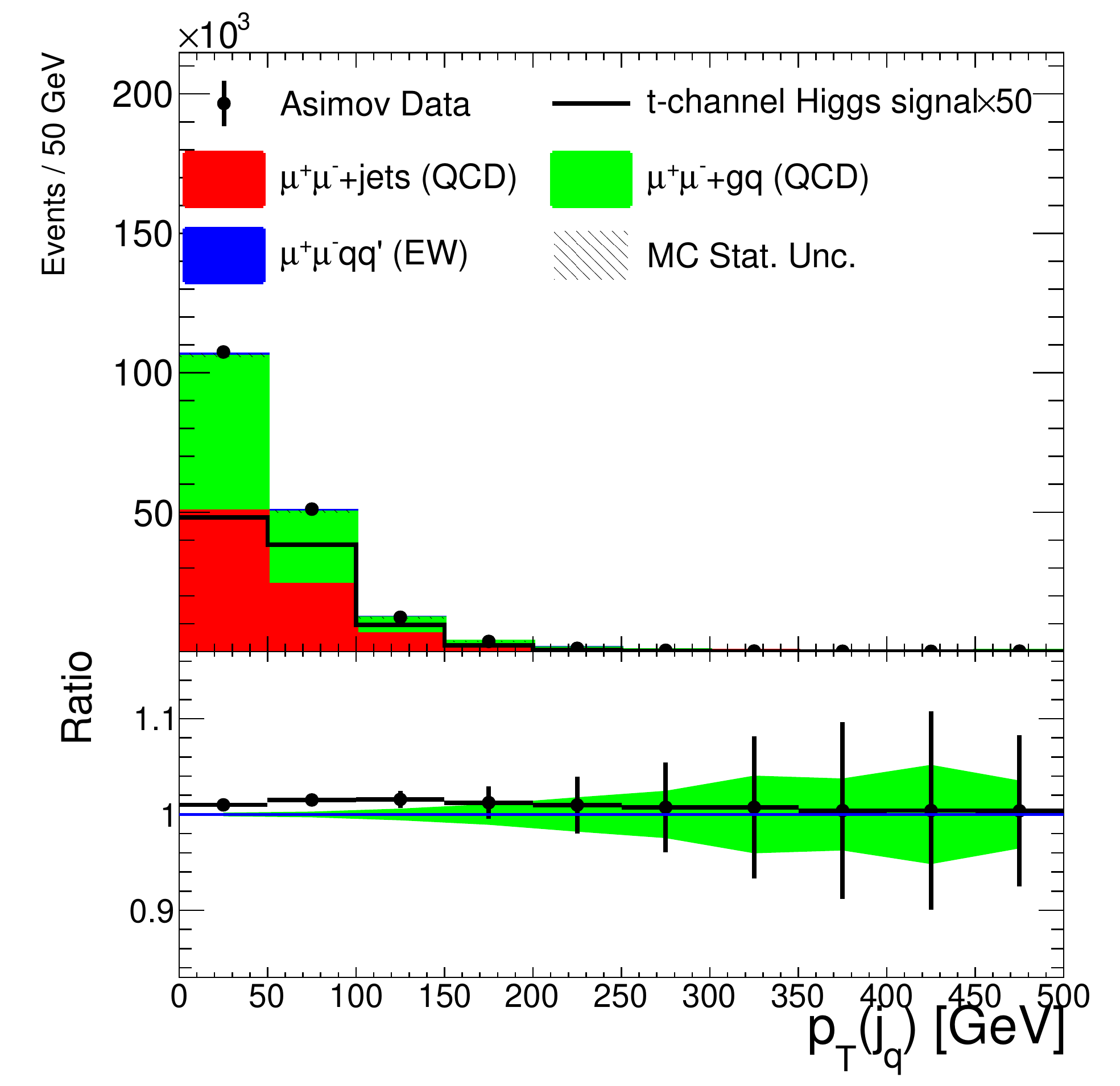}
     \includegraphics[width=0.32\textwidth]{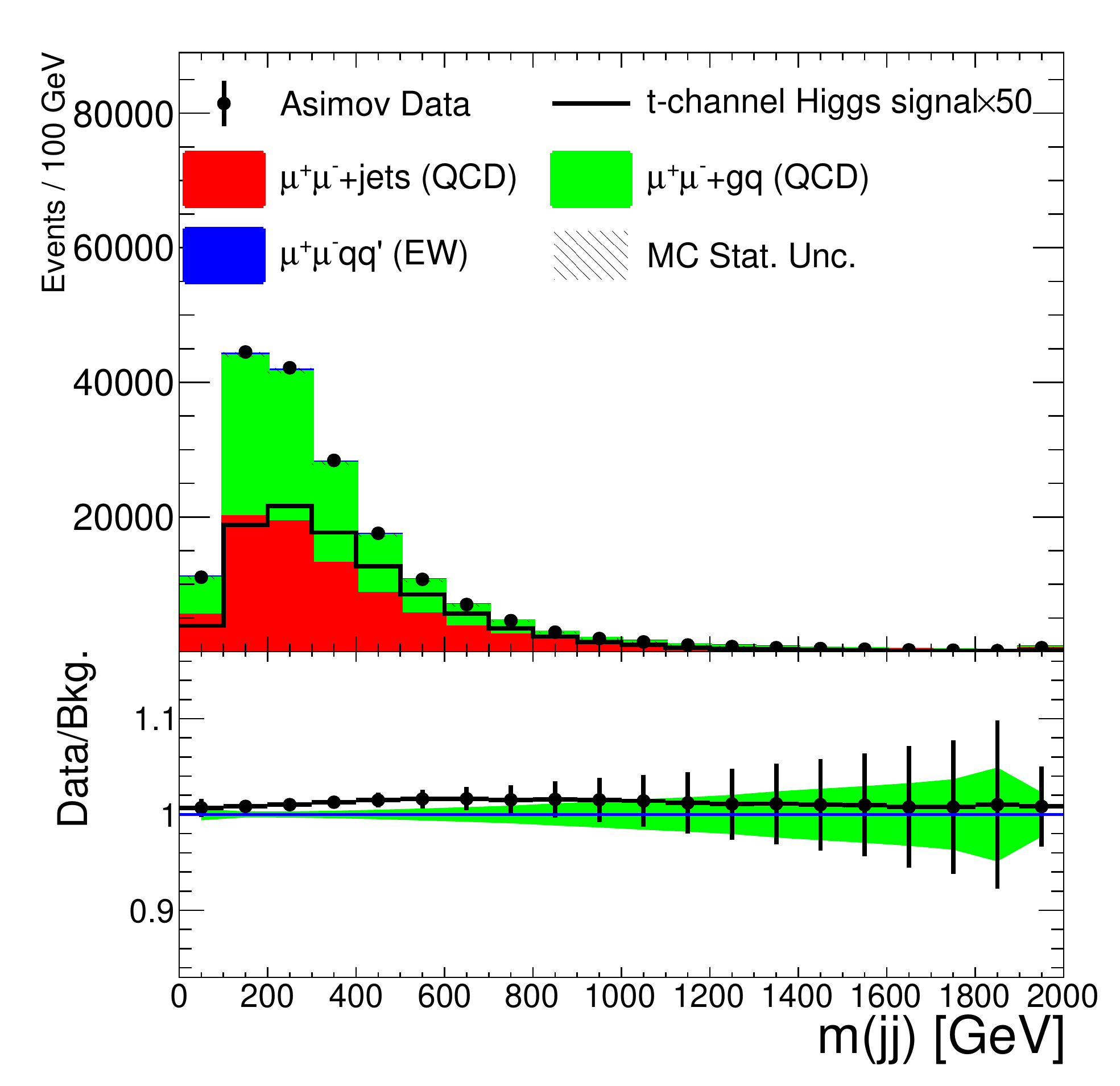}\\
     \includegraphics[width=0.32\textwidth]{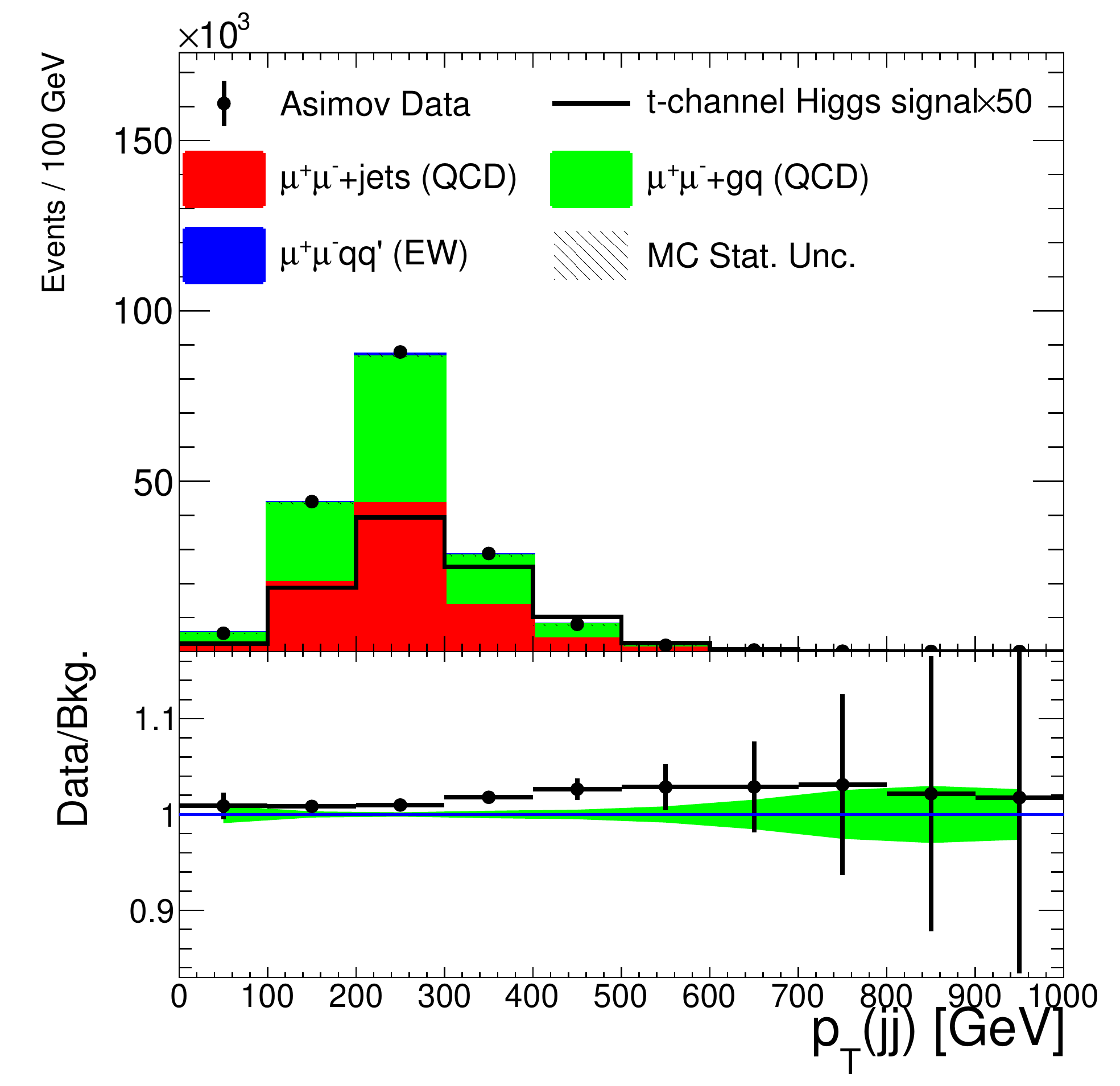}
     \includegraphics[width=0.32\textwidth]{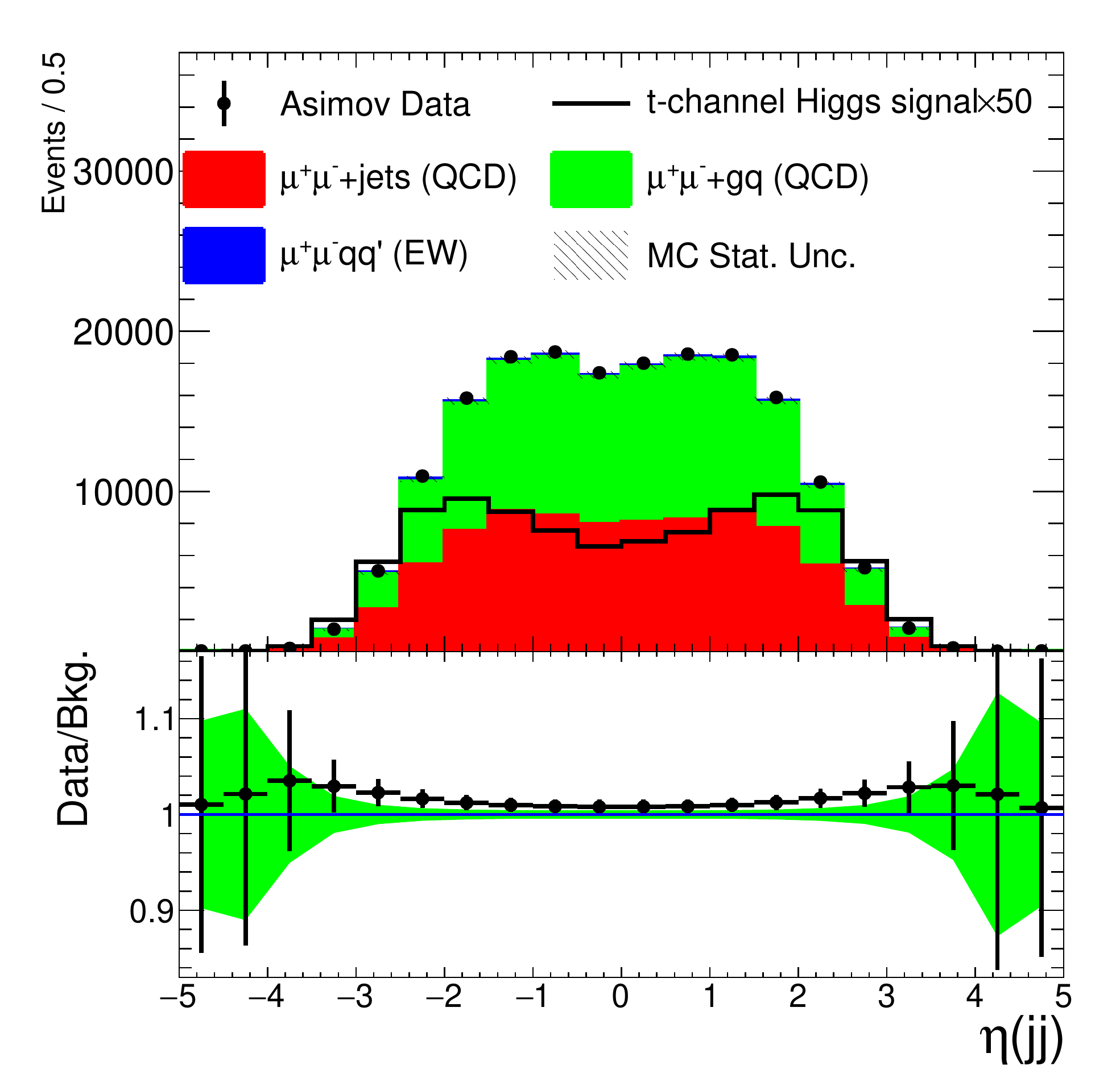}
     \includegraphics[width=0.32\textwidth]{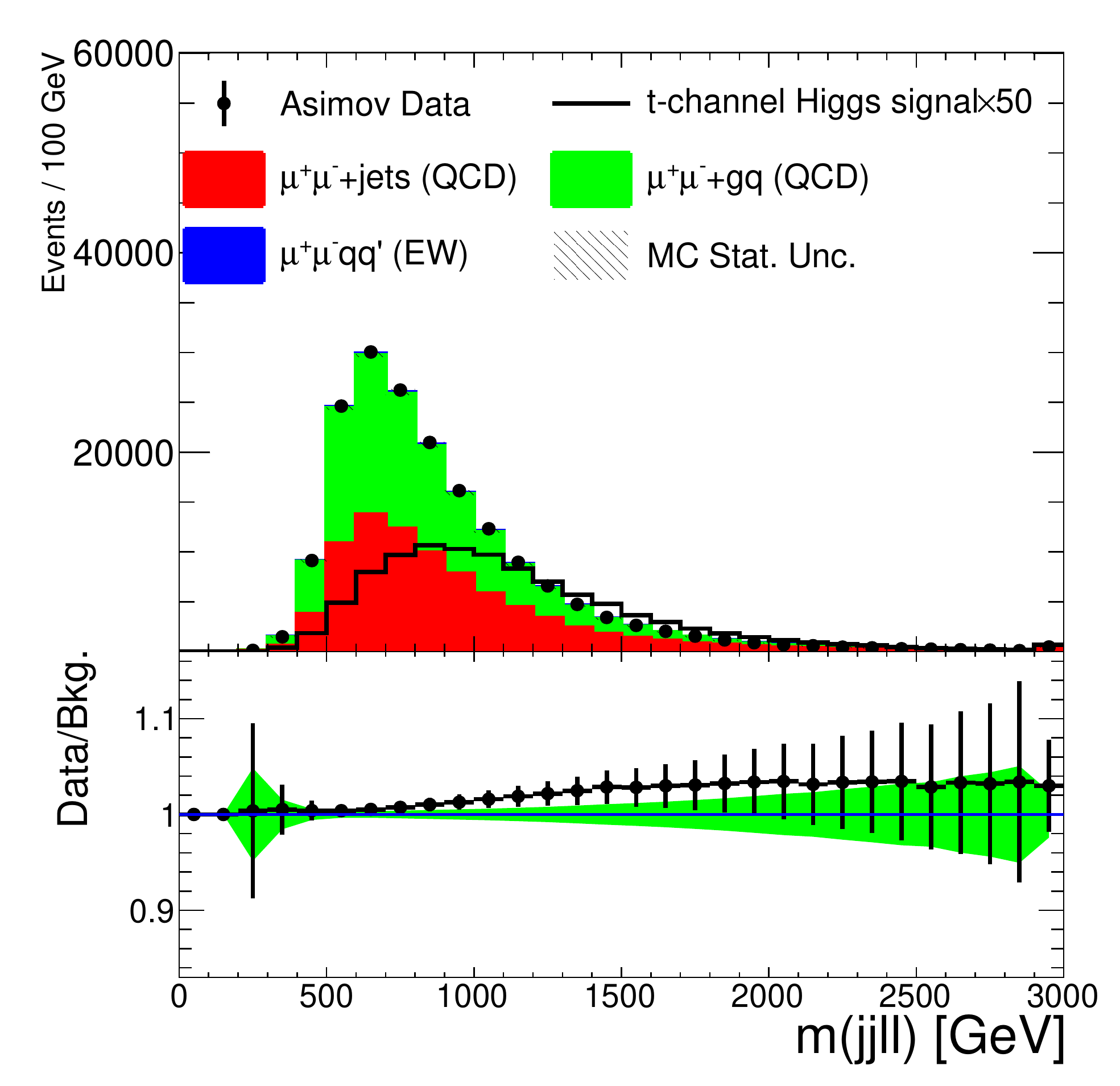}\\
     \includegraphics[width=0.32\textwidth]{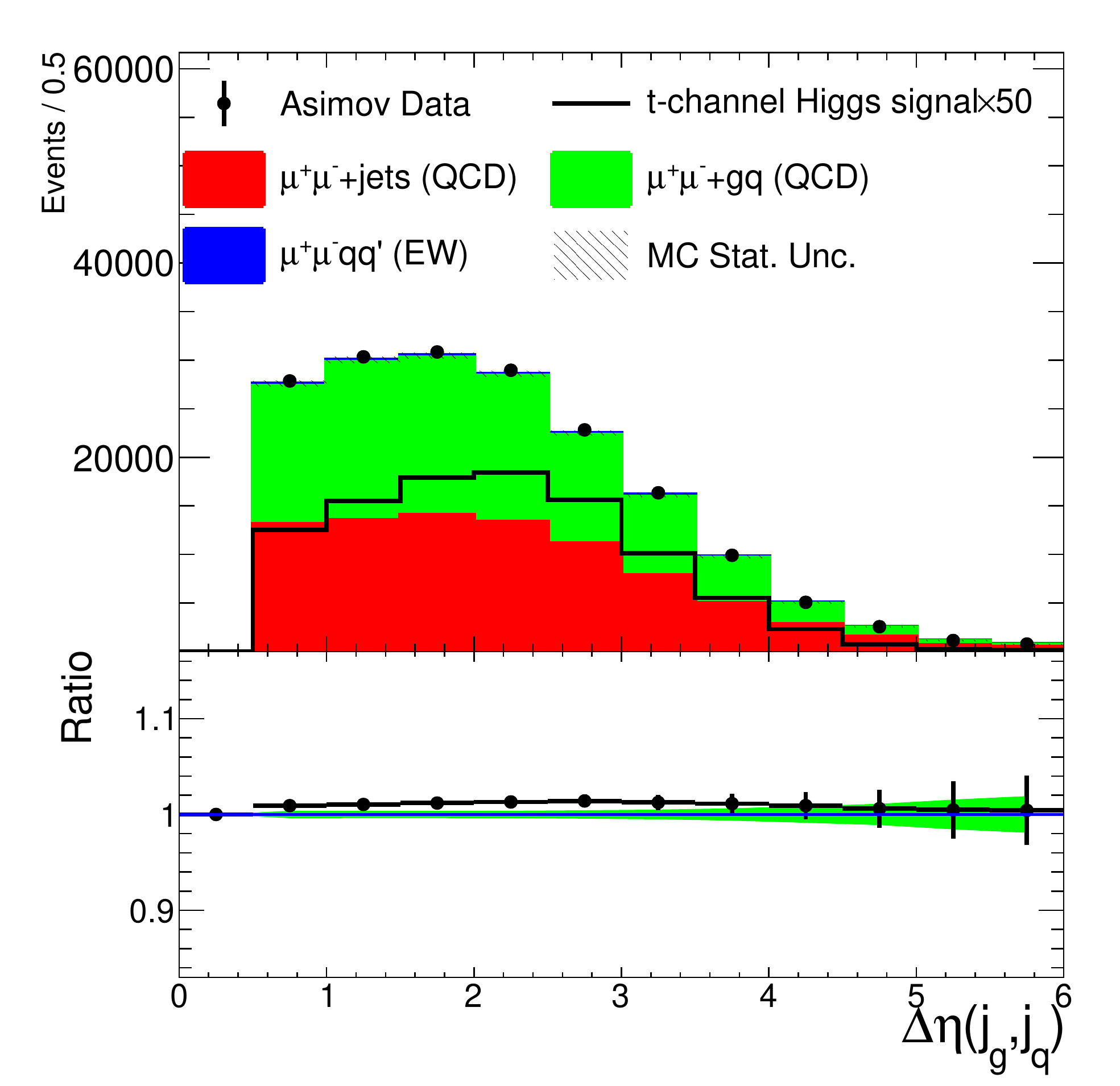}
     \includegraphics[width=0.32\textwidth]{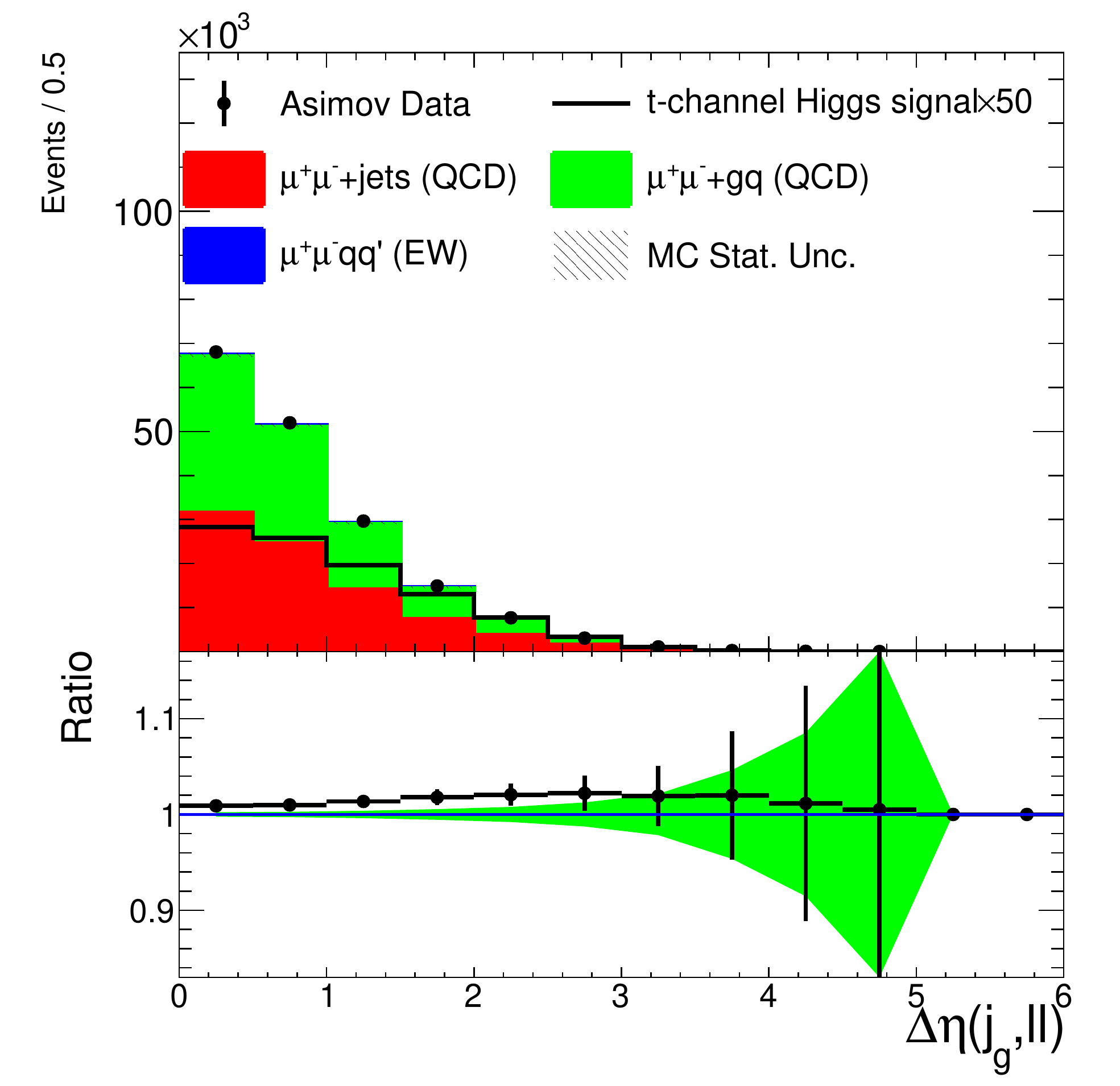}
     \includegraphics[width=0.32\textwidth]{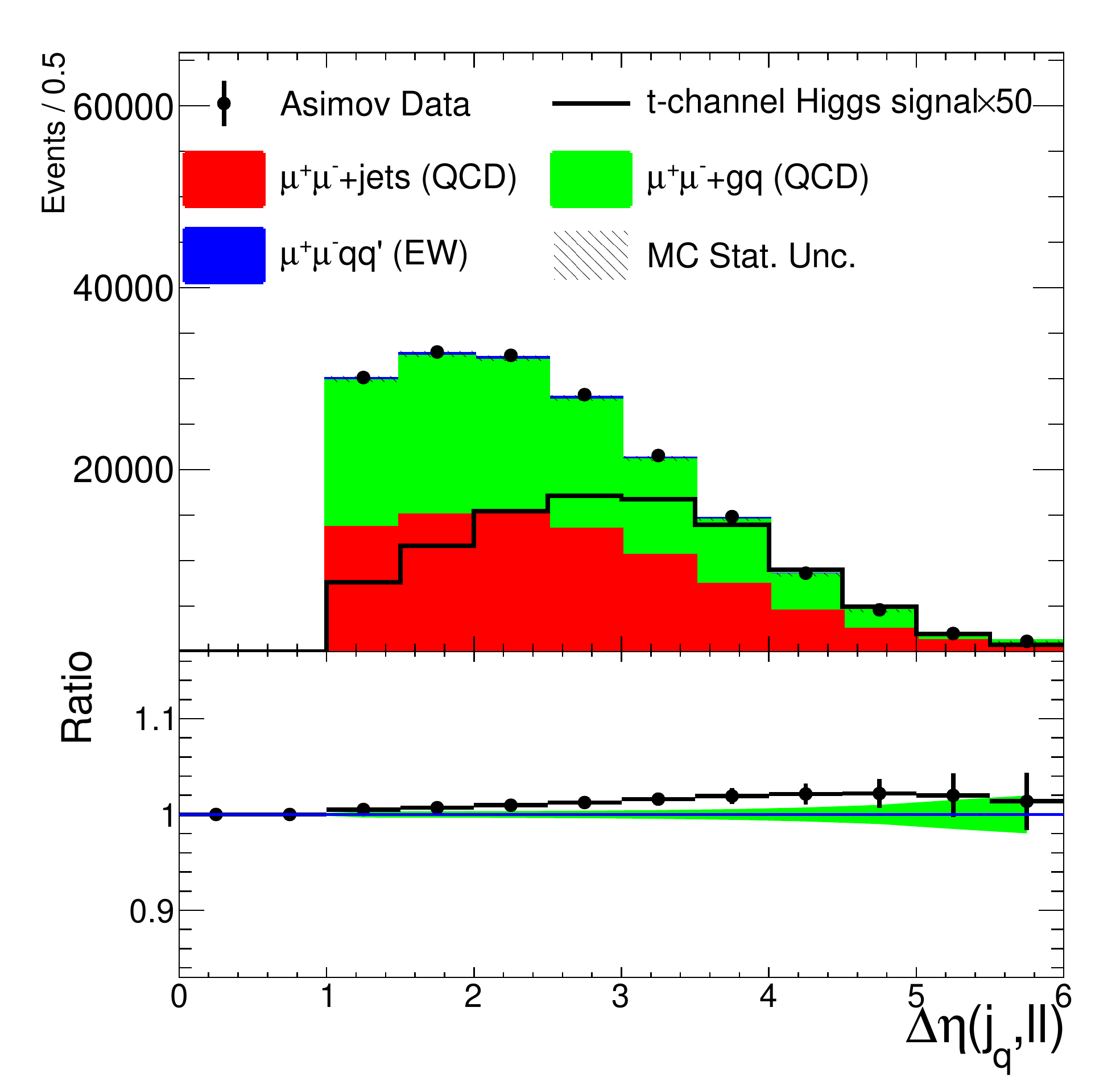}
     \caption{\label{fig:input}
     The distributions of $m(ll)$ (Top Left), $\pt(l_2)$ (Top Right), $|\eta(j_g)-\eta(j_q)|$ (Bottom Left) and $|\eta(j_q)-\eta(ll)|$ (Bottom Right) and the corresponding cuts. The asimov data uses $\kappa=50$. 
     }
 \end{figure}

\begin{figure}[htbp]
     \centering
     \includegraphics[width=0.32\textwidth]{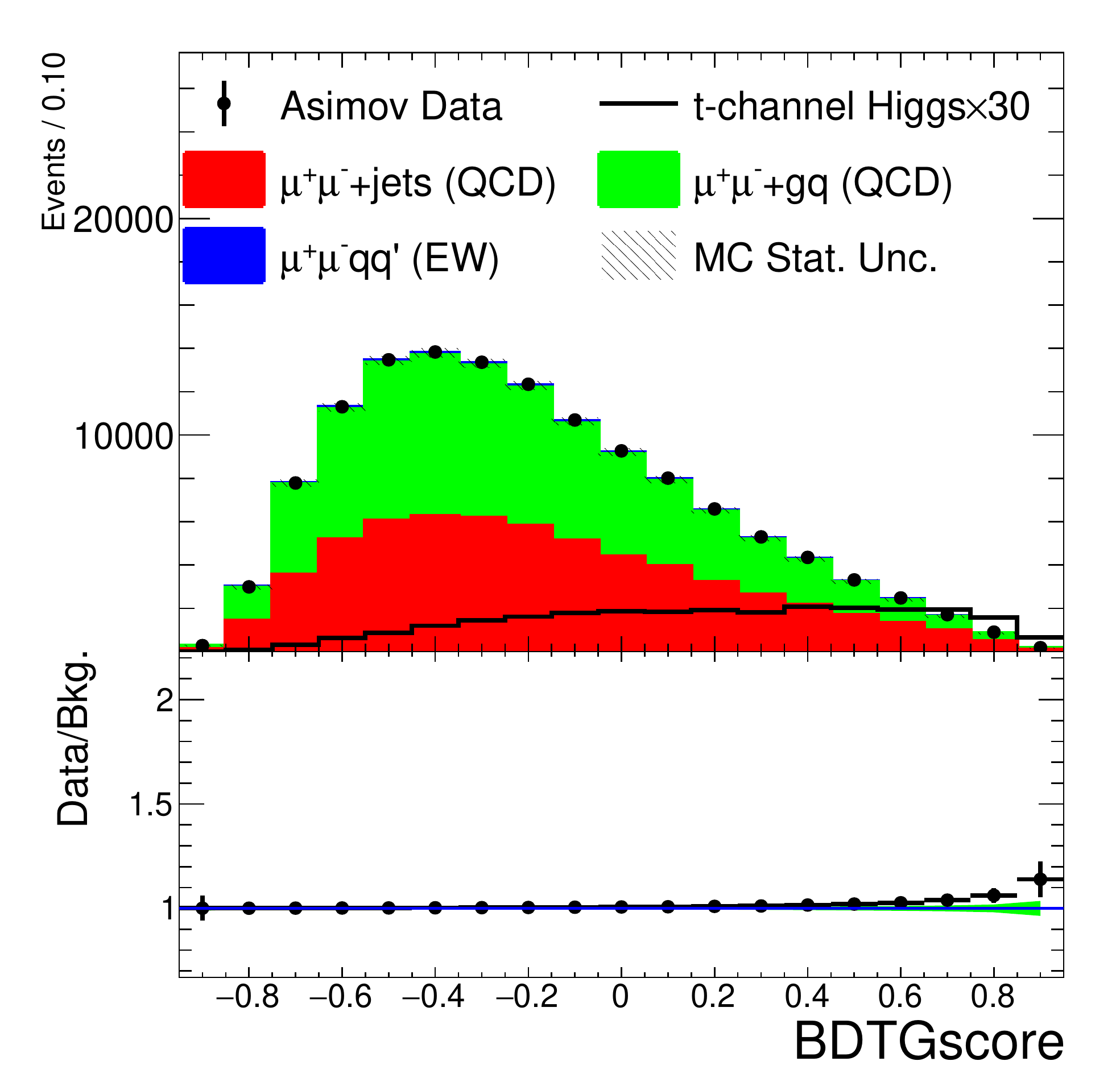}
     \includegraphics[width=0.32\textwidth]{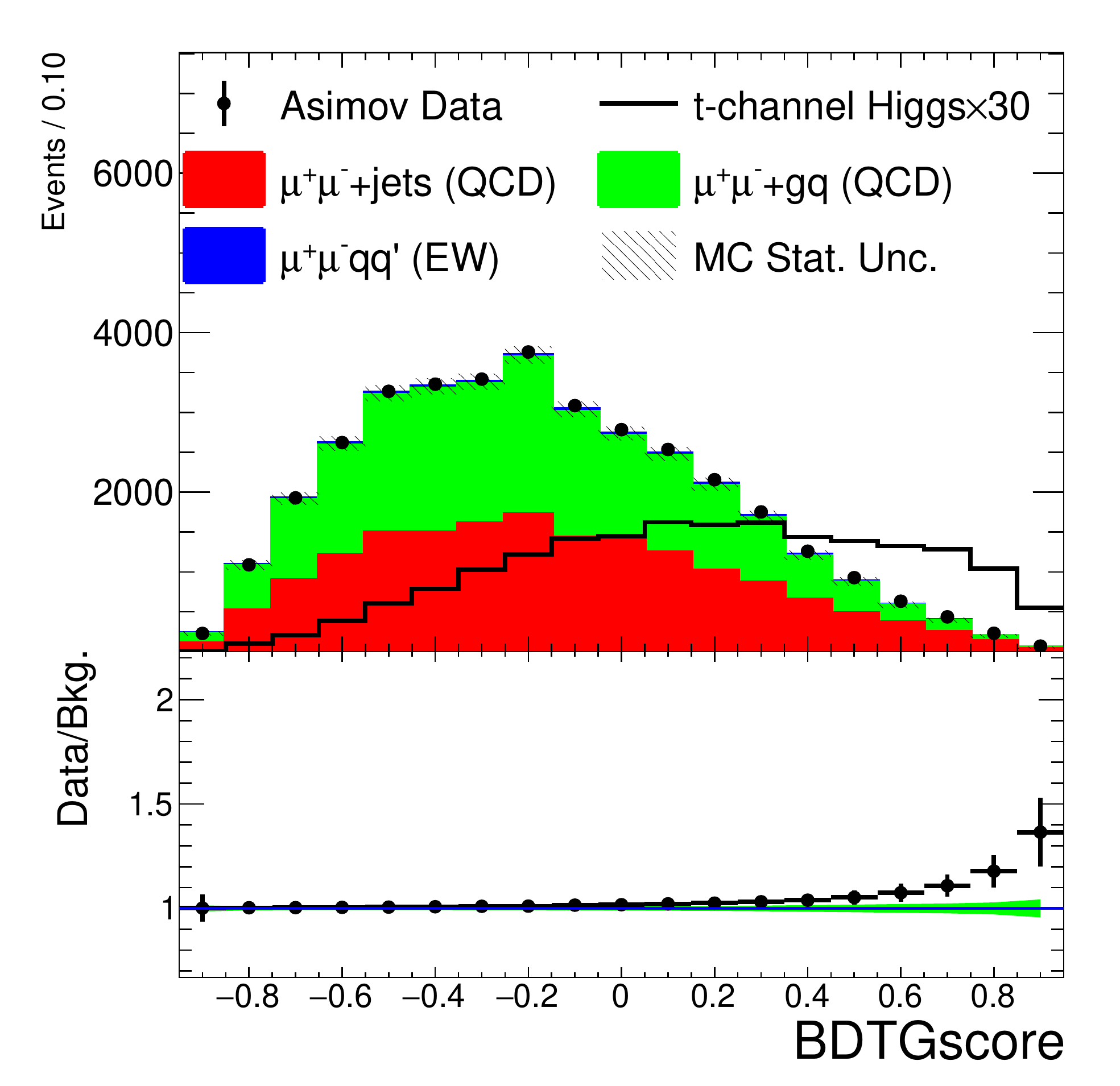}
     \includegraphics[width=0.32\textwidth]{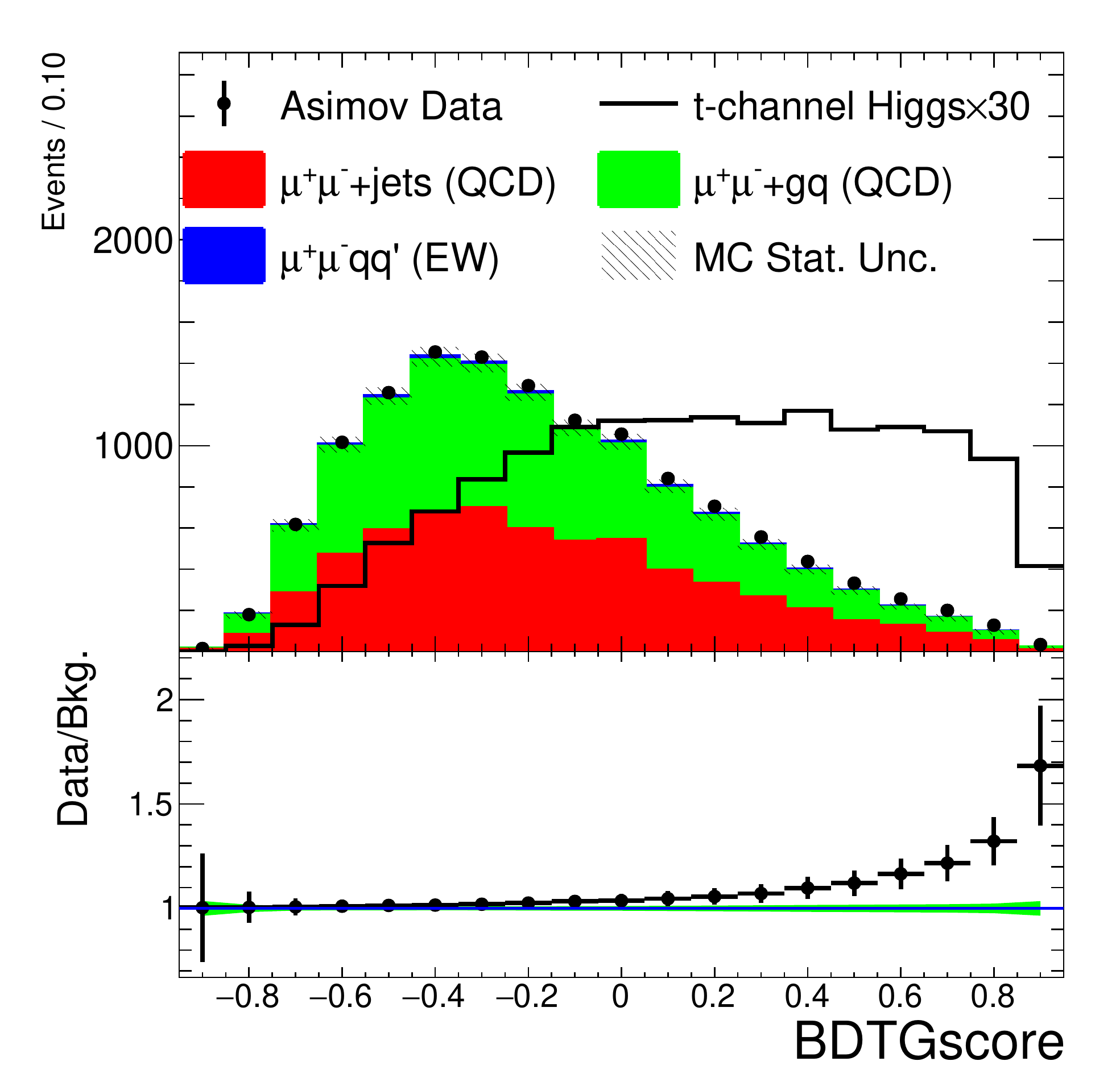}
     \caption{\label{fig:bdtscore}
     The Gradient BDT score distribution in the ``low ptll'' (L), ``middle ptll'' (M) and ``hight ptll''  signal region (R).
     }
 \end{figure}

To further improve the signal sensitivity, we resort to the multi-variate analysis (MVA) and the Gradient BDT alogrithm. We design three signal regions according to $\pt(ll)$: 1) ``low ptll'': $200<\pt(ll)<300$, 2) ``middle ptll'': $300<\pt(ll)<400$~GeV and 3) ``high ptll'': $\pt(ll)>400$~GeV. The training is performed separately in the signal regions. The input variables are shown in Fig.~\ref{fig:input}. The final BDT score distributions are shown in Fig.~\ref{fig:bdtscore}.

\begin{figure}[htbp]
     \centering
     \includegraphics[width=0.45\textwidth]{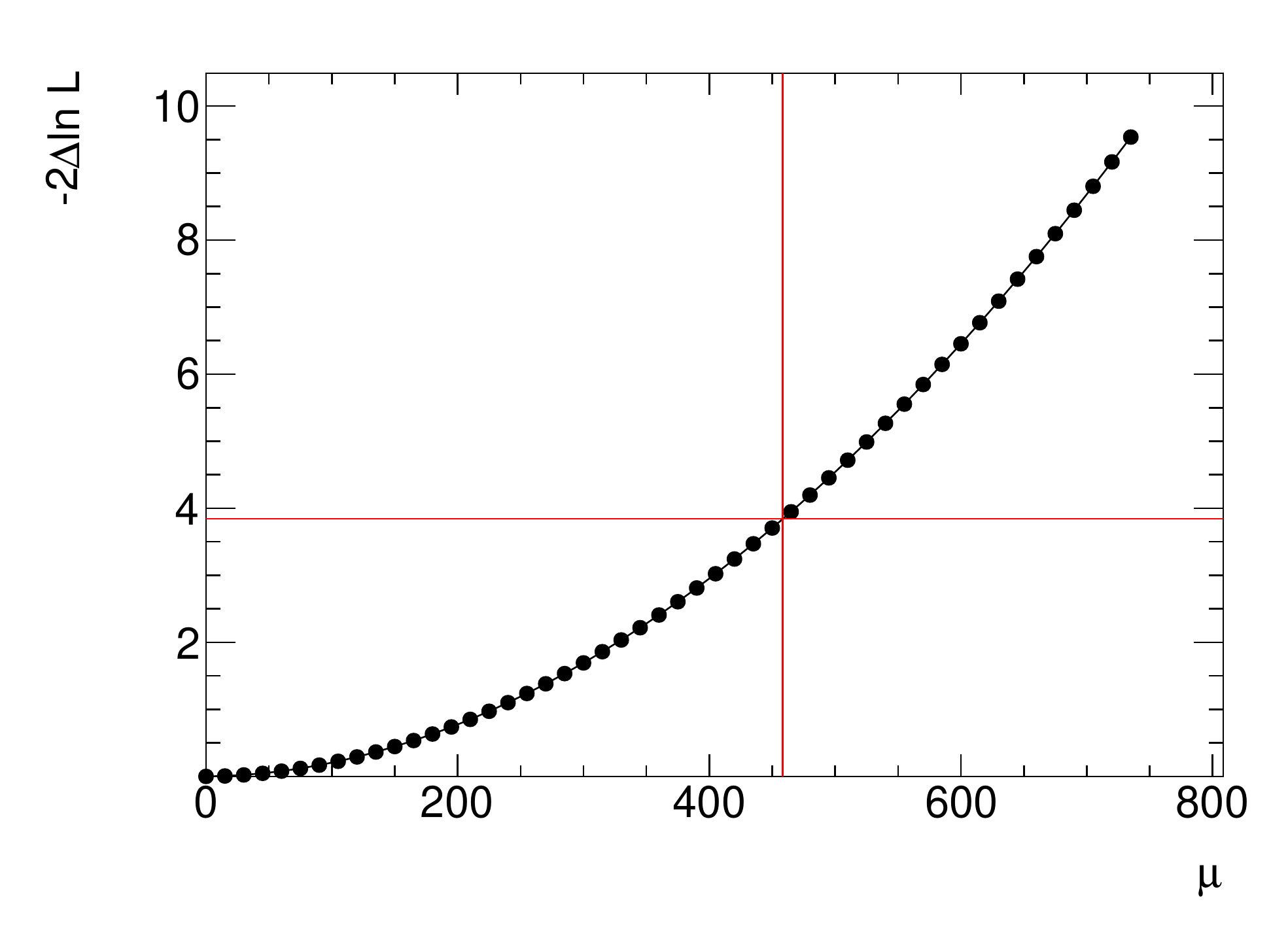}
     \caption{\label{fig:llscan}
     The $-2$ logarithmic likelihood function as a function of the signal strength. The red lines indicate the upper limit at 95~\% Confidence Level of the signal strength.
     }
 \end{figure}

 Taking the BDT scores as observables, the signal sensitivity is obtained by performing likelihood fits. Figure~\ref{fig:llscan} shows the $-2$ logarithmic likelihood function as a function of the signal strength $\mu$. The upper limit at 95~\% confidence level is determined to be about 458. As $\mu \propto \kappa^2$, it is tranlated to be the upper limit of $\kappa$, namely, 21.4. If we assume the sensitivity is similar for $l=e$, the upper limit of $\kappa$ is expected to be
 $\frac{21.4}{\sqrt{\sqrt{2}}}\approx 18$  by combining the two cases $l=e$ and $l=\mu$. By the spirit of the ``off-shell'' method, it can be interpreted as an upper limit of the Higgs width, namely, $\Gamma_H < \frac{458}{\sqrt{2}}\Gamma_H^{\SM}\approx1.3$~GeV. This is similar to the sensitivity in the direct measurement.

 \section{Summary}
 In summary, we propose to study the process $g+q \to g+q+l^++l^-$ ($l=e,\mu$). As the Higgs boson contributes in the $t$ channel, it can be used to measure the off-shell Higgs coupling product $y_gy_Z$. Based on the performance of the ATLAS detector in Run~II and assuming a $p$-$p$ dataset of 140~fb$^{-1}$ at $\sqrt{s}=13$~TeV, the expected upper limit at 95~\% confidence level of the coupling modifier $\kappa$ is about 18. Using the ``off-shell'' method, it is translated to a constraint
 on the Higgs width, namely, $1.3$~GeV. This is similar to the sensitivity in the direct measurement and thus complementary to other measurement methods. It seems best to search for $t$-channel Higgs production in $e+p$ colliders using the process like $e+g\to e + g+l^++l^-$ because the background-to-signal ratio would be much smaller than in hadron colliders.

\section{Acknowledgments}
I would like to thank Fang Dai for her encouragement. This work is supported by the Young Scientists Fund of the National Natural Science Foundation of China (Grant No. 12105140).  


\begin{thebibliography}{99}
    \bibitem{higgs_observation_atlas}
        ATLAS Collaboration, Phys. Lett. B 716 (2012) 1, arXiv:1207.7214
    \bibitem{higgs_observation_cms}
        CMS Collaboration, Phys. Lett. B 716 (2012) 30, arXiv:1207.7235
    \bibitem{higgswidth0}
        N. Kauer and G. Passarino, JHEP 08 (2012) 116 , arXiv:1206.4803.
    \bibitem{higgswidth1}
        F. Caola and K. Melnikov, Phys. Rev. D 88 (2013) 054024.
\end{thebibliography}
\end{document}